\RequirePackage[T1]{fontenc}
\documentclass[12pt]{article}
\usepackage[height=8.85in,width=6.45in]{geometry}

\usepackage[utf8]{inputenc}
\usepackage{amsmath}
\usepackage{amssymb}
\usepackage{mathtools}
\numberwithin{equation}{section}
\usepackage{slashed}
\usepackage{braket}
\usepackage[svgnames]{xcolor}
\usepackage[colorlinks,linktocpage=true,citecolor=DarkGreen,linkcolor=FireBrick]{hyperref}
\usepackage{cite}
\usepackage{graphicx}
\usepackage{tikz}
\usetikzlibrary{decorations.markings}
\usepackage{tikz-cd}
\usepackage{times}
\usepackage[scaled]{couriers}
\usepackage{bm}
\usepackage{subfig}
\usepackage{mathrsfs}
\usepackage{amsthm}

\usepackage{xcolor}
\usepackage{mdframed}

\renewenvironment{figure}[1][]{
  \begin{originalfigure}[#1]
    \begin{mdframed}[linecolor=black!0,backgroundcolor=black!1]
}{
    \end{mdframed}
  \end{originalfigure}
}

\theoremstyle{plain}

\theoremstyle{definition}

\def\d{{\rm d}}
\def\i{{\mathsf i}}

\def\Hom{\mathop{\mathrm{Hom}}}
\usepackage{slashed}

\def\cD{{\cal D}}

\def\cH{{\cal H}}

\def\cN{{\cal N}}

\def\cR{{\cal R}}

\def\cV{{\cal V}}

\def\bA{{\mathbb A}}

\def\bC{{\mathbb C}}

\def\bH{{\mathbb H}}

\def\bQ{{\mathbb Q}}
\def\bR{{\mathbb R}}

\def\bZ{{\mathbb Z}}

\def\sL{{\mathsf L}}
\def\sM{{\mathsf M}}
\def\sN{{\mathsf N}}

\def\sT{{\mathsf T}}

\def\sX{{\mathsf X}}
\def\sY{{\mathsf Y}}

\def\sb{{\mathsf b}}

\def\so{{\mathsf o}}
\def\sp{{\mathsf p}}

\def\su{{\mathsf u}}

\def\SU{\mathrm{SU}}

\def\Sp{\mathrm{Sp}}

\def\Spin{\mathrm{Spin}}

\def\su{\mathfrak{su}}

\def\so{\mathfrak{so}}
\def\sp{\mathfrak{sp}}

\def\g{\mathfrak{g}}

\def\beq#1\eeq{\begin{align}#1\end{align}}

\def\north{\mathrm{north}}
\def\south{\mathrm{south}}

\begin{document}
\hypersetup{pageanchor=false,
  pdftitle={The tertiary invariant and discrete Green--Schwarz couplings in heterotic string theory},
  pdfauthor={Hiroki Wada and Kazuya Yonekura}}

\begin{titlepage}

\begin{flushright}
TU-1331
\end{flushright}

\vskip 3cm

\begin{center}

{\Large \bfseries 
The tertiary invariant and discrete Green--Schwarz couplings in heterotic string theory
}

\vskip 1cm
Hiroki Wada and Kazuya Yonekura
\vskip 1cm

\begin{tabular}{ll}
Department of Physics, Tohoku University, Sendai 980-8578, Japan
\end{tabular}

\vskip 1cm

\end{center}

\noindent
The tertiary invariant is a recently introduced pairing between two-dimensional $\mathcal N=(0,1)$ supersymmetric quantum field theories, motivated by topological modular forms.
We evaluate it directly from its definition for the sigma models on $\mathrm{Sp}(2)$ and $\mathrm{SU}(3)$ with unit $H$-flux, paired with the internal theories of the non-supersymmetric $\mathfrak{so}(16)\oplus\mathfrak{so}(16)$ heterotic string and of the near-horizon region of the non-supersymmetric heterotic 6-brane, respectively.
Regarding the target spaces as $S^3$ bundles over spheres, we fill in every fiber with a four-disk carrying a gauge instanton and reduce the invariant to the index of a Dirac operator on the base.
We find nonzero values of order three and two, which determine the Green--Schwarz couplings of the two theories on $\mathrm{Sp}(2)$ and $\mathrm{SU}(3)$, respectively.
The former, a $\mathbb Z_3$-valued discrete topological term of the $\mathfrak{so}(16)\oplus\mathfrak{so}(16)$ theory, was previously studied through anomaly inflow to NS5-branes. Our computation, which uses only the worldsheet theories, agrees exactly with that result and thus confirms the anomaly inflow.

\end{titlepage}
\hypersetup{pageanchor=true}

\setcounter{tocdepth}{3}


\tableofcontents
\clearpage


\section{Introduction and summary}
\label{sec:introduction}
\interfootnotelinepenalty=10000

Two-dimensional $\cN=(0,1)$ supersymmetric quantum field theories (SQFTs) are the minimally supersymmetric theories in two dimensions, and they are the worldsheet theories of heterotic string theory~\cite{Gross:1984dd,Gross:1985fr,Gross:1985rr}.
Their dynamics can be studied through quantities that are invariant under continuous deformations.
A basic example is the elliptic genus~\cite{Witten:1986bf}.

The search for further invariants has been guided by the Segal--Stolz--Teichner conjecture~\cite{Segal1988,Segal2007,StolzTeichner2004,Stolz:2011zj} that appropriate deformation classes of $\cN=(0,1)$ SQFTs are described by topological modular forms (TMF).
Motivated by this conjecture, several invariants have been studied by using known results on TMF.
One is the mod-2 elliptic genus~\cite{Tachikawa:2023nne}, an analog of the mod-2 index of the Dirac operator.
Another is the secondary invariant~\cite{Gaiotto:2019asa,Gaiotto:2019gef,Yonekura:2022reu}, an SQFT counterpart of the Bunke--Naumann invariant~\cite{MR3278648}.
The ordinary and mod-2 elliptic genera are collectively called the primary invariants.
More broadly, the Segal--Stolz--Teichner conjecture has been discussed in various contexts; see e.g.~\cite{Gukov:2018iiq,Gaiotto:2018ypj,Gukov:2019lzi,Johnson-Freyd:2020itv,Tachikawa:2021mvw,Tachikawa:2021mby,Lin:2021bcp,Lin:2022wpx,Albert:2022gcs,Chae:2022kcq,Berwick-Evans:2023ejh,Saxena:2024eil,Johnson-Freyd:2024rxr,Berwick-Evans:2024nrr,Debray:2024czl,Lin:2024qqk,Bao:2025xhl,Gukov:2025nmk,Tachikawa:2025awi,Pei:2025cud,Tachikawa:2026jaj,Basile:2026trt,Li:2026yjv,Yu:2026zah,Yoshida:2026nfb,Pei:2026vrk,Albert:2026cwy}.

In~\cite{Tachikawa:2025flw}, a further invariant beyond the primary and secondary ones was constructed for theories on which these invariants all vanish, and was accordingly called the tertiary invariant.
It is an SQFT counterpart of the pairing on TMF that follows from its Anderson self-duality~\cite{BrunerRognes,Tachikawa:2023lwf}.
Let $\mathrm{SQFT}_{\nu}(\mathrm{pt})$ be the abelian group of bordism classes of SQFTs with gravitational anomaly $\nu$, and let $\bA_{\nu}$ be its subgroup consisting of the classes whose primary and secondary invariants all vanish.
These notions are reviewed in Section~\ref{sec:review}.
The tertiary invariant is then a bilinear form
\begin{align}
\label{eq:intro-W}
W:\bA_d\times\bA_{-22-d}\longrightarrow\bQ/\bZ.
\end{align}
Here and below, $[\sM]$ denotes the bordism class of an SQFT $\sM$, and we write $W(\sX,\sT)$ for the value of $W$ on $([\sX],[\sT])\in\bA_{d}\times\bA_{-22-d}$.
For such a pair, the degrees add up to $-22$, which is the gravitational anomaly of the matter sector of a worldsheet theory of heterotic string theory.
If $\sX$ is the sigma model with a $d$-dimensional target $X$, then $\sX\times\sT$ is the matter sector of the worldsheet theory of a heterotic string theory on the spacetime $X$ with the internal theory $\sT$.
Let $\Phi_{\sT}(X)$ be the phase of the partition function of this $d$-dimensional spacetime theory on $X$, with the bosonic fields fixed and all gauge fields turned off.
The invariant determines it as
\begin{align}
\label{eq:intro-phase}
\Phi_{\sT}(X)=\exp\bigl[-2\pi\i W(\sX,\sT)\bigr],
\end{align}
as explained in Section~\ref{sec:anomaly-inflow}.
Since the definition of the invariant involves several choices of SQFTs related to $\sX$ and $\sT$, $W(\sX,\sT)$ is not manifestly well-defined.
In~\cite{Tachikawa:2025flw}, it was shown that the tertiary invariant is well-defined under a few assumptions suggested by TMF.
The most crucial one is $\mathrm{SQFT}_{-21}(\mathrm{pt})=0$, i.e.~every SQFT with $\nu=-21$ can be continuously deformed to a theory with spontaneously broken supersymmetry.
This is the SQFT counterpart of the known fact $\mathrm{TMF}_{-21}(\mathrm{pt})=0$, and is closely related to the absence of global anomalies in heterotic string theories~\cite{Tachikawa:2021mby,Yonekura:2022reu}.

When $\sX$ is a sigma model, the theories appearing in the definition of $W$ can be realized geometrically.
The invariant for this case appeared earlier in~\cite{Tachikawa:2024ucm,Kaidi:2024cbx} as a Green--Schwarz coupling in heterotic string theory.
In~\cite{Kaidi:2024cbx} and~\cite{Tachikawa:2025flw}, the invariant was evaluated directly from its definition in the following example: $\sX$ is the sigma model on $S^3\times S^3$ with unit $H$-flux through each factor, and $\sT$ is the chiral fermionic CFT of central charge $14$ containing the $(\mathfrak e_7)_1\times(\mathfrak e_7)_1$ current algebra.
The values for other pairs, including those studied in this paper, were obtained by other methods reviewed below, and their direct evaluations were left for a separate study~\cite{Tachikawa:2025flw}.

The aim of this paper is to carry out such direct evaluations for two pairs.
The first is $([\Sp(2)_{H=1}],[\sT_{16,16}])\in\bA_{10}\times\bA_{-32}$, where $\Sp(2)_{H=1}$ is the sigma model on the group manifold $\Sp(2)$ with unit $H$-flux, and $\sT_{16,16}$ is the chiral fermionic CFT of central charge $16$ whose maximal chiral algebra is $\widehat{\so}(16)_{1}\times\widehat{\so}(16)_{1}$.
The second is $([\SU(3)_{H=1}],[\sT_{16}])\in\bA_{8}\times\bA_{-30}$, where $\SU(3)_{H=1}$ is defined similarly, and $\sT_{16}$ is the chiral fermionic CFT of central charge $15$ whose maximal chiral algebra is $\widehat{\su}(16)_{1}$.
The internal theories $\sT_{16,16}$ and $\sT_{16}$ appear in the classification of chiral fermionic CFTs of central charge at most $16$~\cite{BoyleSmith:2023xkd}.
The theory $\sT_{16,16}$ is the internal theory of the non-supersymmetric $\so(16)\oplus\so(16)$ heterotic string theory~\cite{Alvarez-Gaume:1986ghj,Dixon:1986iz,Kawai:1986vd}.
The key feature of the target spaces $\Sp(2)$ and $\SU(3)$ is that they are $S^3$ bundles over spheres,
\begin{align}\label{eq:intro-fibration}
  S^{3}\to\Sp(2)\to S^{7},\qquad
  S^{3}\to\SU(3)\to S^{5},
\end{align}
with the unit $H$-flux through each $S^3$ fiber.
Supplying the flux by a gauge instanton on a four-disk bounded by the fiber plays a key role in our calculations.
With the conventions specified in the main text, the results of the calculations are summarized as
\begin{align}\label{eq:main-result}
  \boxed{\quad
  \begin{aligned}
  \Phi_{\sT_{16,16}}\bigl(\Sp(2)_{H=1}\bigr)&=\exp\bigl[-2\pi\i W(\Sp(2)_{H=1},\sT_{16,16})\bigr]=\exp\Bigl(\frac{2\pi\i}{3}\Bigr),\\
  \Phi_{\sT_{16}}\bigl(\SU(3)_{H=1}\bigr)&=\exp\bigl[-2\pi\i W(\SU(3)_{H=1},\sT_{16})\bigr]=-1.
  \end{aligned}
  \quad}
\end{align}
Since these phases are nontrivial, none of the theories $\Sp(2)_{H=1}$, $\SU(3)_{H=1}$, $\sT_{16,16}$ and $\sT_{16}$ can be deformed continuously to a theory with spontaneously broken supersymmetry; in other words, their classes are nonzero in $\bA_{10}$, $\bA_{8}$, $\bA_{-32}$ and $\bA_{-30}$, respectively.

The low-energy effective theory of heterotic string theory may contain topological terms.
The target space of heterotic string theory is equipped with a string structure when all gauge fields are turned off.
In this situation, the topological terms are classified by $\Hom(\Omega^{\mathrm{String}}_{d}(\mathrm{pt}),\bR/\bZ)$, where $\Omega^{\mathrm{String}}_{d}(\mathrm{pt})$ is the string bordism group and $d$ is the dimension of the target space.
For $d=10$, it is known that $\Omega^{\mathrm{String}}_{10}(\mathrm{pt})\cong\bZ_{2}\times\bZ_{3}$, and the $\bZ_{3}$ summand is generated by the group manifold $\Sp(2)$ with unit $H$-flux~\cite{MR287553,MR1989190}.
As discussed in~\cite{Tachikawa:2024ucm,Tachikawa:2025flw}, the invariant $W(\Sp(2)_{H=1},\sT_{16,16})$ is related to the $\bZ_{3}$-valued discrete topological term in the non-supersymmetric $\so(16)\oplus\so(16)$ heterotic string theory.

This $\bZ_{3}$-valued term was previously studied through anomaly inflow to NS5-branes~\cite{Tachikawa:2024ucm}.
Consider the two supersymmetric heterotic string theories, with gauge algebras $\so(32)$ and $\mathfrak e_8\oplus\mathfrak e_8$.
When all gauge fields are turned off, they cannot be distinguished perturbatively, since the gravitational parts of their Green--Schwarz couplings agree, but they can still differ by a discrete topological term.
The worldvolume theory of an NS5-brane can have an anomaly of the rotation symmetry of its normal bundle.
Since $\Sp(2)$ is the unit sphere bundle of a four-plane bundle over $S^{7}$, the phase assigned to $\Sp(2)_{H=1}$ gives this anomaly evaluated on $S^{7}$.
In~\cite{Tachikawa:2024ucm}, the difference between these anomalies in the two theories was extracted from the anomaly polynomials of the worldvolume theories.
It was then argued that the two theories differ by a nonzero $\bZ_{3}$-valued topological term, and that the $\so(16)\oplus\so(16)$ string carries the same nonzero term.
This term was revisited in~\cite{Basile:2025mnj}, where it was determined from anomaly inflow and a proposed spectrum of the NS5-brane in the $\so(16)\oplus\so(16)$ string~\cite{Basile:2023knk}.

The invariant $W(\SU(3)_{H=1},\sT_{16})$ was determined in~\cite{Kaidi:2024cbx}, where the non-supersymmetric heterotic 6-brane~\cite{Kaidi:2023tqo} was studied.\footnote{See~\cite{Gibbons:1987ps,Garfinkle:1990qj,Horowitz:1991cd} for the supergravity solution of the 6-brane.}
The near-horizon region of this brane is described by the theory $\sT_{16}$ together with a linear dilaton.
The invariant $W(\SU(3)_{H=1},\sT_{16})$ is the Green--Schwarz coupling of the background with the internal theory $\sT_{16}$ and the sigma model $\SU(3)_{H=1}$.
Its nonzero value implies that $[\sT_{16}]$ is nontrivial, which supports the topological stability of the 6-brane.
Rather than being computed directly, the value was obtained from that of the $S^3\times S^3$ example above, whose internal theory describes the near-horizon region of the non-supersymmetric 4-brane.
This was done by using general properties of the Green--Schwarz coupling and relations among the bordism classes of the internal theories.

Thus, the values in~\cite{Tachikawa:2024ucm,Basile:2025mnj} rely on the worldvolume theories of NS5-branes, while the value in~\cite{Kaidi:2024cbx} is determined from that for another pair.
In contrast, we evaluate the invariant directly from its definition in~\cite{Tachikawa:2025flw} for the pairs themselves, using only the sigma models and the internal theories.
In particular, we determine $W(\Sp(2)_{H=1},\sT_{16,16})$ without reference to any worldvolume theory of NS5-branes; through the anomaly inflow explained in Sections~\ref{sec:anomaly-inflow} and~\ref{sec:comparison}, this in turn determines the global anomaly of the rotation symmetry of the normal bundle of the NS5-brane in the $\so(16)\oplus\so(16)$ string.

The rest of the paper is organized as follows.
\begin{itemize}
\item In Section~\ref{sec:anomaly-inflow}, we explain the physical meaning of the invariant in terms of anomaly inflow to branes. This section provides motivation; apart from the conventions for characteristic classes and orientations fixed there, it is not needed for the computations in later sections.
\item In Section~\ref{sec:review}, we review the notions of SQFTs and the tertiary invariant of~\cite{Tachikawa:2025flw}, together with the coupling of an internal theory to a sigma model. Readers familiar with~\cite{Tachikawa:2025flw} may skip most of this section, except for our conventions in Section~\ref{sec:review-coupling}.
\item In Sections~\ref{sec:disk-bundle}--\ref{sec:reduction}, we set up the calculation for $S^3$ bundles over spheres such as \eqref{eq:intro-fibration}. We realize the theories appearing in the definition of $W$ by filling in every $S^3$ fiber with a four-disk carrying a gauge instanton, and reduce $W$ to a Dirac index on a sphere.
\item In Sections~\ref{sec:sp2} and~\ref{sec:su3}, we carry out the computations for $\mathrm{Sp}(2)_{H=1}$ and $\mathrm{SU}(3)_{H=1}$, respectively, and derive the results \eqref{eq:main-result}.
\item In Section~\ref{sec:comparison}, we compare our calculation for $\mathrm{Sp}(2)_{H=1}$ with that of~\cite{Tachikawa:2024ucm} from the worldvolume theories of NS5-branes. The two calculations agree exactly, which confirms the anomaly inflow explained in Section~\ref{sec:anomaly-inflow}.
\end{itemize}

\paragraph{Note added.}
While this paper was being finalized, we became aware of the independent and closely related work~\cite{Li:2026coordinate}.
We thank the authors of~\cite{Li:2026coordinate} for their helpful comments on a draft of this paper and for coordinating the simultaneous submission to the arXiv.
We also thank Yuji~Tachikawa for bringing this work to our attention.

\section{The physical meaning of the invariant}
\label{sec:anomaly-inflow}
Here we explain the physical meaning of the invariant studied in this paper. It computes the phase $\Phi = Z/|Z|$ of the bulk partition function $Z$.
Anomaly inflow~\cite{Callan:1984sa} relates this phase, evaluated on a sphere bundle linking an
NS-brane, to the anomaly theory of its worldvolume.\footnote{An NS-brane is a magnetic source of the $B$-field, as we define later.} This relation is expected for a general heterotic string with an arbitrary internal worldsheet theory, not only for the supersymmetric ones. We review the perturbative part of anomaly inflow in $d$ dimensions, with the two
supersymmetric heterotic strings in $d=10$, whose gauge algebras are
$\mathfrak{so}(32)$ and $\mathfrak e_8\oplus\mathfrak e_8$, as examples. Then we explain the relation expected between the invariant and a global anomaly of the brane in the general case.

\subsection{The anomaly theory and boundary orientation}
\label{sec:inflow-boundary}

An anomalous boundary theory has a well-defined
partition function if it is combined with a bulk theory. On a closed manifold, this bulk partition function is a phase. These
are the amplitudes of the invertible field theory describing the boundary
anomaly~\cite{Witten:2019bou}. 

The bulk theory in the target space of string theory is gapless, but we may still consider the phase of its partition function for anomaly inflow to the brane.\footnote{It would be interesting to consider the generalization of anomaly inflow when the bulk does not have a mass gap. However, we note that there are no massless fermion modes after the reduction on the three-sphere linking the brane, introduced in Section~\ref{sec:inflow-ns5}.}

Before discussing the details of anomaly inflow, let us set our conventions.
For an oriented $n$-manifold $Y$ and an $(n-1)$-form $\omega$, we give
the boundary the orientation for Stokes' theorem:
\begin{equation}
\label{eq:inflow-stokes}
\int_Y \d\omega=\int_{\partial Y}\omega.
\end{equation}
In oriented coordinates in which the interior is $x^1\leq0$, the outward
normal is $+\partial_{x^1}$ and the boundary has the orientation
$\d x^2\wedge\cdots\wedge \d x^n$. Thus the outward normal precedes an
oriented boundary frame. We use this rule for every boundary below.

For a complex chiral fermion on a $d$-dimensional $X=\partial Y$, with
$\dim Y=d+1$, let $\mathcal D_d^+(X)$ be its chiral Dirac operator.
Write $\eta(Y)$ for the Atiyah--Patodi--Singer (APS) invariant~\cite{Atiyah:1975jf} of the bulk Dirac operator
with APS boundary conditions. The bulk--boundary partition function is
\begin{equation}
\label{eq:inflow-fermion}
\left|\operatorname{Det}\mathcal D_d^+(X)\right|
  \exp\bigl(-2\pi \i\eta(Y)\bigr).
\end{equation}
Throughout this paper, the normalization and sign of $\eta$ are those of
\cite{Tachikawa:2025flw}. To specify them, let a closed $Y$ bound a
$(d+2)$-manifold $Z$ over which the background extends, and let $\mathcal I_{d+2}$
be the fermion anomaly polynomial. The APS index of the Dirac operator
$\mathcal D_{d+2}(Z)$ then obeys
\begin{equation}
\label{eq:inflow-aps}
\begin{aligned}
\operatorname{Ind}\mathcal D_{d+2}(Z)
&=\int_Z \mathcal I_{d+2}+\eta(Y),\\
\exp\bigl(-2\pi \i\eta(Y)\bigr)
&=\exp\left(2\pi \i\int_Z \mathcal I_{d+2}\right).
\end{aligned}
\end{equation}
The anomaly polynomial $\mathcal I_{d+2}$ is defined with this sign. Reality
conditions for Majorana fermions replace determinants by Pfaffians and supply the corresponding
factors of one-half in $\mathcal I_{d+2}$.

To discuss only perturbative anomalies, choose local Chern--Simons forms, denoted $\mathrm{CS}[\mathcal I]$ and normalized by
$\d\,\mathrm{CS}[\mathcal I]=\mathcal I$. The perturbative version of \eqref{eq:inflow-fermion} is
\begin{equation}
\label{eq:inflow-cs}
\operatorname{Det}\mathcal D_d^+(X)\,
\exp\left(2\pi \i\int_Y\mathrm{CS}[\mathcal I_{d+2}]\right).
\end{equation}
The positive
Chern--Simons sign follows from \eqref{eq:inflow-aps}. This local
description retains the perturbative anomaly but need not retain the full
global anomaly.

\subsection{The Green--Schwarz construction}
\label{sec:inflow-ten-dimensional}

This subsection concerns the $d$-dimensional theory on its own. The
spacetime is an arbitrary $d$-manifold $X$ carrying the heterotic
background, and $Y$ is a $(d+1)$-manifold with $\partial Y=X$ over which
that background extends. No NS-brane is assumed
here.

We first fix the characteristic forms. Each of
them is built from the curvature of a chosen connection.
For a real bundle $E$ we write $p_j(E)$ for its Pontryagin forms,
$\chi(E)$ for the Euler form of an oriented bundle of even rank, and
\begin{equation}
\label{eq:inflow-lambda}
\lambda(E)=\frac{p_1(E)}{2}
\end{equation}
for a spin bundle. Several bundles enter at the same time below, so the
argument of $\lambda$ is always displayed in this section. For a gauge algebra with
curvature $F$, we normalize a four-form $c$, which we call the level-one
four-form, on the $\mathfrak{su}(2)$ subalgebra corresponding to a simple root by
\begin{equation}
\label{eq:inflow-level-one}
c=\operatorname{ch}_2
 =\frac12\operatorname{tr}\left(\frac{\i F}{2\pi}\right)^2,
\end{equation}
with the trace in the fundamental doublet representation. For
$\mathfrak{so}(32)$ we write $s_j$ for the Pontryagin forms computed in the
vector representation, so that $c=s_1/2$, and for
$\mathfrak e_8\oplus\mathfrak e_8$ we write $a,a'$ for the level-one forms
of the two factors, so that $c=a+a'$. The gauge-invariant three-form field
strength of the $B$-field is denoted by $H$, and $TX$ is the tangent
bundle of the spacetime. The Bianchi identity for $H$ is
\begin{equation}
\label{eq:inflow-bulk-bianchi}
\d H=c-\lambda(TX).
\end{equation}

Let $\mathcal I_{d+2}$ be the anomaly polynomial
of the $d$-dimensional fields.\footnote{Its sign is fixed as follows. In
an oriented orthonormal frame in $2n$ dimensions we take the chirality matrix as
$\bar\gamma=\i^{-n}\gamma^1\cdots\gamma^{2n}$, and a positive-chirality
fermion coupled to a bundle $E$ has
$\mathcal I^{+}_{2n+2}=[\widehat A\operatorname{ch}(E)]_{2n+2}$, the
degree-$(2n+2)$ part of the product of the A-roof genus of the tangent
bundle and the Chern character. Negative chirality contributes the
opposite sign and Majorana--Weyl fermions a factor $1/2$.} For a general heterotic string, it contains the factor
$c-\lambda(TX)$, the right-hand side of the Bianchi identity
\eqref{eq:inflow-bulk-bianchi}; denote the remaining $(d-2)$-form by $\mathcal J_{d-2}$.
Thus
\begin{equation}
\label{eq:inflow-factorization}
\mathcal I_{d+2}=\bigl(c-\lambda(TX)\bigr)\mathcal J_{d-2}.
\end{equation}
For example, for the two supersymmetric heterotic theories in $d=10$, with gauge
algebra $\mathfrak{so}(32)$ or $\mathfrak e_8\oplus\mathfrak e_8$, the
factors $\mathcal J_8$ in the notation fixed above are
\begin{equation}
\label{eq:inflow-bulk-polynomials}
\begin{aligned}
\mathcal J_8^{\mathfrak{so}(32)}(TX)
&=\frac{3p_1(TX)^2-4p_2(TX)}{192}
  -\frac{\lambda(TX)s_1}{24}
  +\frac{s_1^2-2s_2}{12},\\
\mathcal J_8^{\mathfrak e_8\oplus\mathfrak e_8}(TX)
&=\frac{3p_1(TX)^2-4p_2(TX)}{192}
  -\frac{\lambda(TX)(a+a')}{12}
  +\frac{a^2-aa'+a'^2}{6}.
\end{aligned}
\end{equation}
Here the gauginos and the gravitino have positive chirality and the
dilatino negative chirality.

The phase of the $d$-dimensional partition function is given by the construction of \cite{Tachikawa:2025flw}, which refines the Green--Schwarz mechanism~\cite{Green:1984sg}. On a
$(d+1)$-dimensional extension $Y$ over which the background extends, it
multiplies the $\eta$ factor of \eqref{eq:inflow-fermion} by the exponential
of a $B$-field term, and the phase of the partition function is
\begin{equation}
\label{eq:inflow-spectral-gs}
\Phi(X)=\exp\left(-2\pi \i\eta(Y)-2\pi \i\int_Y H\,\mathcal J_{d-2}\right).
\end{equation}
The $(d+1)$-form $H\,\mathcal J_{d-2}$ is the one whose exterior derivative is
the anomaly polynomial: $\mathcal J_{d-2}$ is a polynomial in curvatures and
hence closed, so $\d(H\,\mathcal J_{d-2})=\mathcal I_{d+2}$ by
\eqref{eq:inflow-bulk-bianchi} and \eqref{eq:inflow-factorization}. By
\eqref{eq:inflow-cs}, the perturbative representative of
\eqref{eq:inflow-spectral-gs} has the exponent
$2\pi \i\int_Y(\mathrm{CS}[\mathcal I_{d+2}]-H\mathcal J_{d-2})$. Choose
$\mathrm{CS}[\mathcal I_{d+2}]=\d H\,\mathrm{CS}[\mathcal J_{d-2}]$. Since $H$ has degree three,
\begin{equation}
\label{eq:inflow-transgression}
\begin{aligned}
\mathrm{CS}[\mathcal I_{d+2}]-H\mathcal J_{d-2}
&=\d H\,\mathrm{CS}[\mathcal J_{d-2}]-H\,\d\,\mathrm{CS}[\mathcal J_{d-2}]\\
&=\d\bigl(H\,\mathrm{CS}[\mathcal J_{d-2}]\bigr).
\end{aligned}
\end{equation}
Thus Stokes' theorem gives a coupling $H\,\mathrm{CS}[\mathcal J_{d-2}]$
on $X$. This local coupling is the bulk ingredient of the
inflow argument below.

\subsection{Perturbative inflow for the NS-brane}
\label{sec:inflow-ns5}

We call a magnetic source of the three-form $H$ an NS-brane. It is linked
by a three-sphere and has codimension four. We do not call it an NS5-brane
because, for a general internal worldsheet theory, the spacetime
dimension $d$ need not be ten, and the worldvolume dimension $d-4$ is six
only for $d=10$.

Let $\Sigma\subset X$ be the oriented $(d-4)$-dimensional worldvolume of an
NS-brane, with oriented normal four-plane bundle $V$. A tubular
neighborhood of the brane is the disk bundle $D(V)\to\Sigma$, whose total
space is oriented by fiber followed by base, as is
every bundle total space in this paper: for an $n_1$-dimensional fiber and an $n_2$-dimensional base, the total space is locally oriented as $\mathbb{R}^{n_1+n_2} = \mathbb{R}^{n_1} \times \mathbb{R}^{n_2}$. Each linking sphere is oriented as $S^3=\partial D^4$ in
the convention \eqref{eq:inflow-stokes}. Along the brane, the tangent
bundle of $X$ splits as $TX|_\Sigma=V\oplus T\Sigma$. For a form $\alpha$
on the tubular neighborhood, write $\alpha|_\Sigma$ for its restriction to
the brane, that is, to the zero section of $D(V)\to\Sigma$.
In the tubular neighborhood, we choose the connections on the tangent and gauge bundles of $X$ to be pulled back from connections on $\Sigma$ along the projection $D(V)\to\Sigma$.
Then the restrictions of the characteristic forms $c$ and $\lambda(TX)$ to the tubular neighborhood are pulled back from $\Sigma$.
In particular, these forms on the tubular neighborhood integrate to zero over each fiber $D^4$, and they are the pullbacks of their restrictions to $\Sigma$.

The two simple summands of the transverse rotation algebra
$\mathfrak{so}(4)$ are denoted $\mathfrak{su}(2)_L$ and
$\mathfrak{su}(2)_R$, with $\mathfrak{su}(2)_L$ generated by self-dual elements in $\mathfrak{so}(4)$. Write $c(L),c(R)$ for the
level-one forms as above. In the same orientation,
\begin{equation}
\label{eq:inflow-normal-classes}
\lambda(V)=c(L)+c(R),\qquad
\chi(V)=c(L)-c(R).
\end{equation}

The brane is a magnetic source for $H$: it carries one unit of flux
through a linking three-sphere. We keep this source in the Bianchi
identity as a delta-function four-form localized on the brane.
Let $\delta_\Sigma$ be a delta-function four-form supported on the location of the brane, with unit integral over each fiber disk,
\begin{equation}
\label{eq:inflow-source}
\int_{D^4}\delta_\Sigma=1.
\end{equation}
Such a form (or more precisely its slight smearing) is called a Thom form of $V$, that is, a de Rham representative of the
Poincar\'e dual of the brane.
In the presence of the brane, the Bianchi identity \eqref{eq:inflow-bulk-bianchi} acquires this source,
\begin{equation}
\label{eq:inflow-bianchi-source}
\d H=c-\lambda(TX)+\delta_\Sigma.
\end{equation}
Away from the brane the source drops out.

Integrating \eqref{eq:inflow-bianchi-source} over a fiber disk, the terms
$c-\lambda(TX)$ give nothing, and Stokes' theorem \eqref{eq:inflow-stokes}
leaves the linking flux
\begin{equation}
\label{eq:inflow-flux}
\begin{aligned}
\int_{S^3}H&=\int_{D^4}\d H
=\int_{D^4}\delta_\Sigma=1.
\end{aligned}
\end{equation}
Restricting instead to the brane, a Thom form represents the Euler class
of $V$, and with the adapted connection it is the Euler form,\footnote{This is shown explicitly for a small instanton in Section~\ref{sec:disk-bundle-general-base}.}
\begin{equation}
\label{eq:inflow-delta-euler}
\delta_\Sigma|_\Sigma=\chi(V).
\end{equation}
This is the self-intersection of the brane: the positive unit source
leaves $\chi(V)$ behind on its own worldvolume, with the sign fixed by the
transverse orientation above. Restricting
\eqref{eq:inflow-bianchi-source} to the brane therefore gives the
worldvolume Bianchi identity
\begin{equation}
\label{eq:inflow-bianchi}
\begin{aligned}
\d\bigl(H|_\Sigma\bigr)&=c-\lambda(TX)+\chi(V)\\
&=c-\lambda(T\Sigma)-2c(R) \\
&=:\mathcal K_4,
\end{aligned}
\end{equation}
where the second line uses $\lambda(TX)=\lambda(T\Sigma)+\lambda(V)$ and
\eqref{eq:inflow-normal-classes}.

We now reduce the bulk coupling $H\,\mathrm{CS}[\mathcal J_{d-2}]$
of \eqref{eq:inflow-transgression} near the brane. Let $r$ be the radial
distance from the brane. Reducing the
exterior $r>0$ on its linking three-spheres gives a $(d-3)$-dimensional base
$B$ whose boundary is the brane at $r=0$. The spacetime is
$X=S(V)\to B$ where $B$ includes the radial direction. With the orientations fixed above, 
\begin{equation}
\label{eq:inflow-radial-orientation}
\begin{aligned}
\operatorname{vol}_{d}
&=\d r\wedge\operatorname{vol}_{S^3}\wedge\operatorname{vol}_\Sigma\\
&=\operatorname{vol}_{S^3}\wedge
  \bigl(-\d r\wedge\operatorname{vol}_\Sigma\bigr).
\end{aligned}
\end{equation}
Hence $\operatorname{vol}_B=-\d r\wedge\operatorname{vol}_\Sigma$; the outward
normal $-\partial_r$ at the brane gives $\partial B=\Sigma$ in our Stokes
convention.
The positive flux $\int_{S^3}H=1$ then gives
\begin{equation}
\label{eq:inflow-reduced-bulk}
\exp\left(2\pi \i\int_X H\,\mathrm{CS}[\mathcal J_{d-2}]\right)
=\exp\left(2\pi \i\int_B\mathrm{CS}[\mathcal J_{d-2}]\right).
\end{equation}
These are local representatives; other choices of connections introduce
the corresponding local terms which are not important for the topological classification of anomalies.

As examples, consider the two supersymmetric theories in $d=10$, where the NS-brane is the NS5-brane. In each theory, write $\mathcal I_8$ for the anomaly polynomial of the
NS5-brane worldvolume theory and $\mathcal J_8(TX)$ for the bulk eight-form
of the same theory, given in \eqref{eq:inflow-bulk-polynomials}. Their
difference is divisible by the worldvolume Bianchi four-form $\mathcal K_4$ of \eqref{eq:inflow-bianchi};
writing the remaining four-form as $\mathcal L_4$,
\begin{equation}
\label{eq:inflow-brane-quotient}
\mathcal I_8-\mathcal J_8(TX)=\mathcal K_4\mathcal L_4 .
\end{equation}

For the $\mathfrak{so}(32)$ NS5-brane, $\mathcal I_8$ is the polynomial
$\mathcal I_8^{\mathfrak{sp}(1)}$ of the worldvolume theory with its
dynamical $\mathfrak{sp}(1)$ field, whose level-one class is denoted
$c_{\rm g}$.\footnote{$\mathcal I_8^{\mathfrak{sp}(1)}$, including its sign, is computed as follows. Realize the positive source as a gauge instanton, self-dual in
the transverse orientation. The gaugino variation
$\delta\chi\propto F_{mn}\Gamma^{mn}\epsilon$ then vanishes for $\epsilon$
of negative transverse chirality, which is an $\mathfrak{su}(2)_L$ singlet
and an $\mathfrak{su}(2)_R$ doublet; hence the R-symmetry is
$\mathfrak{su}(2)_R$. With the transverse-before-brane orientation, the
preserved six-dimensional supercharges have negative chirality, hyperinos
positive and gauginos negative. With $\mathbf{32}$ the flavor vector,
$\mathbf2_{\rm g},\mathbf3_{\rm g}$ the fundamental and adjoint of
$\mathfrak{sp}(1)$, and $\mathbf2_L,\mathbf2_R$ the rotation doublets, the
half-hypermultiplet in $(\mathbf{32},\mathbf2_{\rm g})$, the position
hypermultiplet and the vector multiplet~\cite{Witten:1995gx} give
$\mathcal I_8^{\mathfrak{sp}(1)}=\frac12[\widehat A(T\Sigma)
(\operatorname{ch}(\mathbf{32})\operatorname{ch}(\mathbf2_{\rm g})
+\operatorname{ch}(\mathbf2_L)-\operatorname{ch}(\mathbf3_{\rm g})
\operatorname{ch}(\mathbf2_R))]_8$.}
For an NS5-brane associated with the first $\mathfrak e_8$, $\mathcal I_8$ is
the rank-one E-string polynomial $\mathcal I_8^{\mathrm{E\text{-}str}}$,
including its free position hypermultiplet, and $\mathcal K_4$ uses
$c=a+a'$.\footnote{We take $\mathcal I_8^{\mathrm{E\text{-}str}}$ from \cite{Ohmori:2014pca} and fix its signs. With $\mathcal I_4'=[p_1(V)+p_1(T\Sigma)]/4-a$ and
$\mathcal I_8'=\frac1{48}[p_2(V)+p_2(T\Sigma)-(p_1(V)
-p_1(T\Sigma))^2/4]$ there,
$\mathcal I_8^{\mathrm{E\text{-}str}}=p_2(V)/6-\chi(V)\mathcal I_4'/2
+\mathcal I_4'^2/2-\mathcal I_8'$. The overall sign follows from the
positive chirality of the $\mathfrak e_7$-charged hyperinos on the Higgs
branch, and the minus sign of $\chi(V)\mathcal I_4'/2$ in our normal
orientation from the absence of an $a\,c(L)$ term in that
$\mathfrak{su}(2)_L$-neutral sector. With $p_2(V)=\chi(V)^2$ it gives the second line of
\eqref{eq:inflow-brane-four-forms}.} The two four-forms are\footnote{The signs of $\mathcal I_8^{\mathfrak{sp}(1)}$ and
$\mathcal I_8^{\mathrm{E\text{-}str}}$ are fixed by the previous two
footnotes, not by requiring the inflow identities to hold.}
\begin{equation}
\label{eq:inflow-brane-four-forms}
\begin{aligned}
\mathcal L_4^{\mathfrak{so}(32)}
&=c_{\rm g}+\frac{\lambda(V)-\lambda(T\Sigma)}{12},\\
\mathcal L_4^{\mathfrak e_8\oplus\mathfrak e_8}
&=\frac{\lambda(V)-\lambda(T\Sigma)}{12}
  -\frac{c(R)}3+\frac{2a-a'}6.
\end{aligned}
\end{equation}

For the worldvolume Green--Schwarz mechanism we add
$-H\mathcal L_4$ to its seven-dimensional anomaly theory. The phase of its perturbative
partition function is therefore 
\begin{equation}
\label{eq:inflow-completed-brane}
\begin{aligned}
&\exp\left[2\pi \i\int_B
\bigl(\mathrm{CS}[\mathcal I_8]-H\mathcal L_4\bigr)\right],\\
&\d\bigl(\mathrm{CS}[\mathcal I_8]-H\mathcal L_4\bigr)=\mathcal J_8(TX).
\end{aligned}
\end{equation}
More explicitly, write the two-form $B$-field as $B_2$, so that
$H=\d B_2+\mathrm{CS}[\mathcal K_4]$. Choosing
$\mathrm{CS}[\mathcal I_8]=\mathrm{CS}[\mathcal J_8]+\mathrm{CS}[\mathcal K_4]\mathcal L_4$ gives
\begin{equation}
\label{eq:inflow-local-gs}
\mathrm{CS}[\mathcal I_8]-H\mathcal L_4
=\mathrm{CS}[\mathcal J_8]-\d(B_2\mathcal L_4).
\end{equation}
The exact term is the local six-dimensional Green--Schwarz coupling
on $\Sigma=\partial B$. The remaining Chern--Simons
amplitude has exactly the form in \eqref{eq:inflow-reduced-bulk}, including the sign.
This establishes perturbative inflow in each theory: the reduced bulk
supplies the anomaly theory that pairs with the brane partition function.

\subsection{The bulk phase and the anomaly theory of the brane}
\label{sec:inflow-global}

Recall that the exterior of the NS-brane in Section~\ref{sec:inflow-ns5} is
$X=S(V)\to B$ with $\partial B=\Sigma$. Write $Z$ for the bulk partition function on $S(V)$ over a closed $B$ with fixed bosonic backgrounds, and $\Phi$ for its phase,
\begin{equation}
\label{eq:inflow-modulus-phase}
Z\bigl(S(V)\bigr)
=\bigl|Z\bigl(S(V)\bigr)\bigr|\,\Phi\bigl(S(V)\bigr).
\end{equation}
Anomaly inflow is the statement that this phase is the partition function
of the anomaly theory of the worldvolume theory of the NS-brane, evaluated on
the $(d-3)$-dimensional base $B$ on which the linking spheres are fibered. Denoting this anomaly by
$\mathcal A_\text{brane}(B)\in\mathbb R/\mathbb Z$, including its worldvolume
Green--Schwarz term,
\begin{equation}
\label{eq:inflow-expected-global}
\Phi\bigl(S(V)\bigr)
\;\stackrel{\mathrm{inflow}}{=}\;
\exp\bigl[2\pi \i\mathcal A_\text{brane}(B)\bigr].
\end{equation}
Both sides are amplitudes on the same base $B$, the left-hand
side after reducing on the positive linking sphere. 

The previous subsection verified the perturbative part of \eqref{eq:inflow-expected-global} for $\mathfrak{so}(32)$ and $\mathfrak e_8\oplus\mathfrak e_8$:
the reduced bulk amplitude \eqref{eq:inflow-reduced-bulk} is exactly the
Chern--Simons amplitude \eqref{eq:inflow-completed-brane} of the
worldvolume theory including the worldvolume Green--Schwarz term. A Chern--Simons description fixes only that
perturbative part, and a nontrivial global anomaly can still exist~\cite{Tachikawa:2024ucm}.

The left-hand side of \eqref{eq:inflow-expected-global} is what the
invariant studied in this paper computes under special conditions. Let $\mathsf{T}$ denote the internal worldsheet
theory and $\mathsf{X}$ the sigma model with target $X$. We consider the partition function of the theory $\mathsf{X} \times \mathsf{T}$ without introducing any gauge bundle. The invariant $W(\mathsf{X},\mathsf{T})$ which we will compute is related to the phase of the bulk partition function $\Phi_\mathsf{T}(X)$ as \cite{Tachikawa:2025flw}
\begin{equation}
\label{eq:inflow-physical-phase}
\Phi_\mathsf{T}(X)=\exp\bigl[-2\pi \i W(\mathsf{X},\mathsf{T})\bigr].
\end{equation}
We will compute it in the situation where $X=S(V)$ and $B$ is a sphere.

\section{Preliminaries on SQFTs}
\label{sec:review}
In this section, we review the tertiary invariant of SQFTs introduced in~\cite{Tachikawa:2025flw}; none of the material here is new.
We start in Section~\ref{sec:review-sqft} by collecting basic notions of two-dimensional $\cN=(0,1)$ SQFTs, including theories with boundary and their APS index.
The tertiary invariant is defined for pairs of bordism classes of SQFTs whose primary and secondary invariants vanish.
These notions are recalled in Section~\ref{sec:review-invariants}, and the definition of the tertiary invariant itself is reviewed in Section~\ref{sec:review-pairing}.
Finally, in Section~\ref{sec:review-coupling}, we explain how an internal theory is coupled to a sigma model, since the resulting theories appear in our construction in Sections~\ref{sec:disk-bundle} and~\ref{sec:bordism}.

\subsection{Basic notions of SQFTs}
\label{sec:review-sqft}

Following~\cite{Tachikawa:2025flw}, we begin by recalling several notions of $\cN=(0,1)$ SQFTs.

\smallskip
\noindent\textbf{Gravitational anomaly.}
The gravitational anomaly of a fermionic quantum field theory in two dimensions is specified by an integer $\nu$, in the sense that its anomaly polynomial is given by
\begin{align}
\label{eq:review-grav}
    -\frac{\nu}{48}\,p_{1},
\end{align}
where $p_{1}$ is the first Pontryagin class.
In this convention, a right-moving Majorana--Weyl fermion has $\nu=+1$ and a left-moving one has $\nu=-1$.
An $\cN=(0,1)$ sigma model with a $d$-dimensional target space contains $d$ right-moving Majorana--Weyl fermions as the superpartners of the bosonic fields, and therefore has $\nu=d$.
For an SQFT with conformal symmetry, $\nu=2(c_R-c_L)$, where $c_R$ and $c_L$ are the right- and left-moving central charges.

Throughout this paper, we consider the Hilbert space on a circle of circumference $2\pi$ with the periodic (Ramond) spin structure.
We denote the Hamiltonian, the momentum operator and the supercharge by $H$, $P$ and $Q$, and set $H_{L}=(H+P)/2$ and $H_{R}=(H-P)/2$.
The supersymmetry algebra is given by $Q^2=H_{R}$.
For a theory with gravitational anomaly $\nu$, the eigenvalues of $P$ lie in
\begin{align}
    P\in\bZ-\frac{\nu}{24}.
\end{align}
A state annihilated by $Q$ has $H_{R}=0$, and hence $H_{L}=P$.

\smallskip
\noindent\textbf{Compact SQFTs.}
A theory is compact if the Hamiltonian $H$ has a discrete spectrum and $e^{-\beta H}$ is trace class, i.e.~$\operatorname{Tr}e^{-\beta H}<\infty$, for every $\beta>0$.
Sigma models on closed manifolds and the internal theories used in our examples are compact.

\smallskip
\noindent\textbf{Orientation reversal.}
The orientation reversal $\overline{\sM}$ of an SQFT $\sM$ is obtained by stacking the Arf theory on $\sM$.
For a sigma model, this corresponds to reversing the orientation of the target space.

\smallskip
\noindent\textbf{Product of SQFTs.}
Let $\sM_{1}$ and $\sM_{2}$ be theories with gravitational anomalies $\nu_{1}$ and $\nu_{2}$.
Their product $\sM_{1}\times\sM_{2}$ is the theory whose Hilbert space is the tensor product of those of $\sM_{1}$ and $\sM_{2}$, and its gravitational anomaly is $\nu_{1}+\nu_{2}$.

\smallskip
\noindent\textbf{Sum of SQFTs.}
For two theories $\sM_{1}$ and $\sM_{2}$ with the same gravitational anomaly, the disjoint union $\sM_{1}\sqcup\sM_{2}$ is the theory whose Hilbert space is the direct sum of those of $\sM_{1}$ and $\sM_{2}$.
It has two superselection sectors, described by $\sM_{1}$ and $\sM_{2}$, respectively.

\smallskip
\noindent\textbf{Theories with boundary.}
Consider first a sigma model whose target space is a manifold $N$ with boundary $\partial N=M$.
We attach the semi-infinite cylinder $\bR_{\geq0}\times M$ to $N$ along $M$, where $\bR_{\geq0}=[0,\infty)$.
Following~\cite{Tachikawa:2025flw}, we call $M$ the boundary of the resulting noncompact theory, although, strictly speaking, its target space has no boundary.

More generally, an SQFT $\sN$ with boundary $\partial\sN=\sM$, where $\sM$ is compact, is a mildly noncompact theory: its continuous spectrum is the same as that of $\bR_{\geq0}\times\sM$.
The coordinate of $\bR_{\geq0}$ and its right-moving superpartner increase the gravitational anomaly by one.
Hence, if the gravitational anomaly of $\sM$ is $\nu_{\sM}$, that of $\sN$ is given by $\nu_{\sN}=\nu_{\sM}+1$.

\smallskip
\noindent\textbf{Gluing of SQFTs.}
Two theories $\sN_{1}$ and $\sN_{2}$ with the same boundary $\sM$ can be glued along $\sM$ by connecting their noncompact regions through a long cylinder $[-a,a]\times\sM$ with $a\gg1$.
We denote the resulting compact theory by $\sN_{1}\cup\overline{\sN_{2}}$.

\smallskip
\noindent\textbf{The APS index.}
Recall first the case of a Dirac operator on a manifold with boundary.
We assume that the Dirac operator on the boundary has no zero modes.
Then, the index under the APS boundary condition equals the number of normalizable zero modes of positive chirality minus that of negative chirality.
These zero modes are localized in the interior region.

Let $\sL$ be an SQFT with boundary $\partial\sL=\sN$ whose gravitational anomaly $\nu_{\sL}$ is even.
Regarding the supercharge $Q$ and the fermion parity $(-1)^F$ as analogs of the Dirac operator and the chirality operator, we define the APS index of $\sL$ as follows.
For simplicity, we assume that $\sN$ has no states annihilated by $Q$.
We decompose the Hilbert space of $\sL$ into eigenspaces of $P$, whose eigenvalues are $n-\nu_{\sL}/24$ with $n\in\bZ$.
In the eigenspace with $P=n-\nu_{\sL}/24$, let $\mathrm{Index}_{n-\nu_{\sL}/24}$ be the number of states annihilated by $Q$ with $(-1)^{F}=+1$ minus the number of those with $(-1)^{F}=-1$.
The APS index of $\sL$ is then
\begin{align}\label{eq:review-aps}
    I_{\sL}&=\kappa\sum_{n\in\bZ}q^{n-\nu_{\sL}/24}\mathrm{Index}_{n-\nu_{\sL}/24},&
    q&=e^{2\pi\i\tau},
\end{align}
where $\operatorname{Im}\tau>0$, and
\begin{align}
    \kappa=
    \begin{cases}
    1,&\nu_{\sL}\equiv0\mod8,\\[2pt]
    \frac12,&\nu_{\sL}\equiv4\mod8.
    \end{cases}
\end{align}
When $\nu_{\sL}\equiv4\mod8$, every $\mathrm{Index}_{n-\nu_{\sL}/24}$ is even because of the Kramers degeneracy, and the factor $\kappa=1/2$ makes the coefficients integers.
The APS index vanishes automatically for $\nu_{\sL}\equiv2\mod4$, and is not defined for odd $\nu_{\sL}$, since $(-1)^{F}$ is not defined.

For two theories $\sL$ and $\sL'$ with the same boundary, the APS index satisfies the gluing law $I_{\sL'\cup\overline{\sL}}=I_{\sL'}-I_{\sL}$.

\smallskip
\noindent\textbf{SQFTs with a corner.}
Let $\sN_1$ and $\sN_2$ be theories with a common boundary, $\partial\sN_1=\partial\sN_2=\sM$.
An SQFT $\sL$ with a corner has two noncompact regions, $\bR_{\geq0}\times\sN_1$ and $\overline{\bR'_{\geq0}\times\sN_2}$, which overlap in $\bR_{\geq0}\times\bR'_{\geq0}\times\sM$.
After a modification of its behavior at infinity, $\sL$ can be regarded as a theory with boundary $\partial\sL=\sN_1\cup\overline{\sN_2}$.
As long as $\sN_1\cup\overline{\sN_2}$ has no zero modes, this modification does not affect the APS index of $\sL$.

\subsection{Primary and secondary invariants}
\label{sec:review-invariants}

We now recall the primary and secondary invariants of SQFTs.
Since they are defined on bordism classes, we first recall bordism of SQFTs.

\smallskip
\noindent\textbf{Bordism classes of SQFTs.}
Two compact SQFTs $\sM_{1}$ and $\sM_{2}$ with the same gravitational anomaly are bordant if $\sM_{1}\sqcup\overline{\sM_{2}}$ is the boundary of a noncompact SQFT.
If a theory is the boundary of a noncompact theory, one can continuously deform it to a theory with spontaneously broken supersymmetry~\cite{Gaiotto:2019asa,Gaiotto:2019gef,Yonekura:2022reu,Tachikawa:2025flw}.
The bordism classes of compact theories with gravitational anomaly $\nu$ form an abelian group $\mathrm{SQFT}_{\nu}(\mathrm{pt})$, in which the sum is given by the disjoint union, the inverse by the orientation reversal, and the class of a boundary is zero.
The product of theories makes $\bigoplus_{\nu}\mathrm{SQFT}_\nu(\mathrm{pt})$ a graded ring.

A bordism invariant is a function on compact SQFTs that depends only on the bordism class.
The invariants below are also additive, i.e.~they define homomorphisms from $\mathrm{SQFT}_\nu(\mathrm{pt})$.

\smallskip
\noindent\textbf{Primary invariants.}
The primary invariants of SQFTs are the ordinary elliptic genus and the mod-2 elliptic genus.
For a compact theory, the ordinary elliptic genus~\cite{Witten:1986bf} is given by the right-hand side of \eqref{eq:review-aps} without the factor $\kappa$.
As explained below \eqref{eq:review-aps}, it can be nonzero only when $\nu\equiv0,4\mod 8$.

One can also consider the mod-2 elliptic genus~\cite{Tachikawa:2023nne}, which can be nonzero only when $\nu\equiv1,2\mod 8$.
When $\nu\equiv1\mod8$, its coefficient in each eigenspace of $P$ is the number of states annihilated by $Q$ modulo two.
When $\nu\equiv2\mod8$, this number is always even, and half of it modulo two is used instead.
Like the ordinary elliptic genus, it is invariant under continuous deformations.

\smallskip
\noindent\textbf{Secondary invariant.}
The secondary invariant~\cite{Gaiotto:2019gef,Yonekura:2022reu} is an SQFT analog of the Bunke--Naumann invariant~\cite{MR3278648} of TMF.
The tertiary invariant is defined only for theories whose secondary invariant vanishes.
This condition is relevant only when $\nu\equiv3\mod4$, which does not occur in our examples, and we do not need the details of the secondary invariant below.

\subsection{Tertiary invariant of SQFTs}
\label{sec:review-pairing}
We now review the tertiary invariant introduced in~\cite{Tachikawa:2025flw}, which is defined on bordism classes whose primary and secondary invariants vanish.
Let $\bA_{\nu}\subset\mathrm{SQFT}_{\nu}(\mathrm{pt})$ denote the subgroup of such classes.
For an integer $d$, the tertiary invariant is a bilinear form
\begin{align}
    W(\bullet,\bullet):\bA_{d}\times\bA_{-22-d}\to\bQ/\bZ.
\end{align}
Its definition relies on the following three conjectures:
\begin{enumerate}
\item
$\mathrm{SQFT}_{-21}(\mathrm{pt})=0$.
\item
$\bA_{\nu}$ is a torsion group.
\item
$[\sX][\sT]=0$ in $\mathrm{SQFT}_{-22}(\mathrm{pt})$ for all $([\sX],[\sT])\in\bA_{d}\times\bA_{-22-d}$.
\end{enumerate}
These conjectures are suggested by the corresponding properties of TMF together with the Segal--Stolz--Teichner conjecture~\cite{StolzTeichner2004,Stolz:2011zj}.
As emphasized in~\cite{Tachikawa:2025flw}, even if the second and third conjectures fail, the tertiary invariant can still be defined on the subset of pairs in $\bA_{d}\times\bA_{-22-d}$ for which these two conjectures hold.

\begin{figure}[htbp]
\centering
\begin{tikzpicture}[scale=2,line width=1pt,font=\small]
  \def\r{1.5}
  \draw (0,0) arc[start angle=180, end angle=270, radius=\r];
  \draw[FireBrick,line width=1.6pt] (0,0) -- (\r,0);
  \draw[SteelBlue,line width=1.6pt] (\r,0) -- (\r,-\r);
  \fill (\r,0) circle (1.4pt);
  \node at (\r/2,-0.5*\r) {$\sL$};
  \node[FireBrick] at (\r/2,0.17) {$\sN^{\sqcup k}$};
  \node[SteelBlue,anchor=west] at (\r+0.08,-0.5*\r) {$\overline{\sY_{0}\times \sT}$};
  \node[anchor=south west] at (\r+0.04,0.06) {$(\sX\times \sT)^{\sqcup k}$};
\end{tikzpicture}
\space 
\caption{The theory $\sL$ with a corner used in the definition \eqref{eq:review-3}. Its two boundary components are $k$ copies of the theory $\sN$ with $\partial\sN=\sX\times \sT$ and the orientation reversal of the product $\sY_{0}\times \sT$ with $\partial \sY_{0}=\sX^{\sqcup k}$. They meet at the corner $(\sX\times \sT)^{\sqcup k}$.}
\label{fig:corner}
\end{figure}
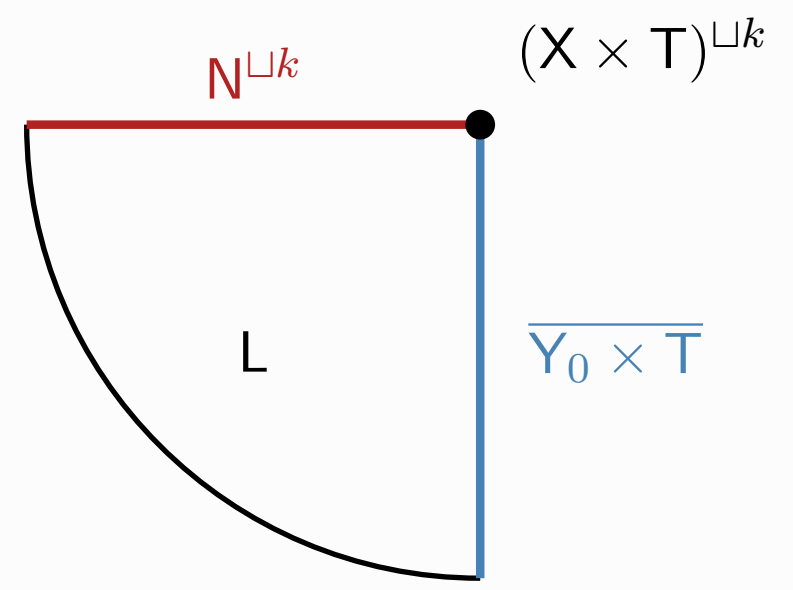

The tertiary invariant is defined as follows~\cite{Tachikawa:2025flw}.
Let $\sX$ and $\sT$ be compact theories whose classes satisfy $[\sX]\in\bA_d$ and $[\sT]\in\bA_{-22-d}$.
By the third conjecture, there is a noncompact theory $\sN$ with $\partial\sN=\sX\times\sT$.
By the second conjecture, $k[\sX]=0$ for some positive integer $k$, i.e.~there is a noncompact theory $\sY_{0}$ with $\partial\sY_{0}=\sX^{\sqcup k}$.\footnote{When $d\equiv3\mod4$, the pair $(k,\sY_{0})$ must satisfy an additional condition involving the secondary invariant~\cite{Tachikawa:2025flw}. This condition plays no role here, since $d=10$ and $8$ in our examples.}
Both $\sN^{\sqcup k}$ and $\sY_{0}\times\sT$ have boundary $(\sX\times\sT)^{\sqcup k}$, and gluing them gives a compact theory $\sN^{\sqcup k}\cup\overline{\sY_{0}\times\sT}$ with gravitational anomaly $-21$.
By the first conjecture, its class in $\mathrm{SQFT}_{-21}(\mathrm{pt})$ vanishes, i.e.~there is a noncompact theory $\sL$ with
\begin{align}
\label{eq:review-2}
    \partial\sL=\sN^{\sqcup k}\cup\overline{\sY_{0}\times\sT},
\end{align}
which we regard as a theory with a corner, as in Figure~\ref{fig:corner}.

Since $\nu_{\sL}=-20\equiv4\mod 8$, we have $\kappa=1/2$ in \eqref{eq:review-aps}, and $I_{\sL}$ is a series in $q^{n+20/24}$ with $n\in\bZ$ and integer coefficients.
Multiplying it by $\eta_{\mathrm D}(\tau)^{4}$, where $\eta_{\mathrm D}(\tau)=q^{1/24}\prod_{m\geq1}(1-q^m)$ is the Dedekind eta function, we obtain a Laurent series in $q$ with integer coefficients.
The tertiary invariant is then defined as
\begin{align}
\label{eq:review-3}
    W(\sX,\sT)=\frac{1}{k}
    \left[
    \eta_{\mathrm D}(\tau)^{4}I_{\sL}(\tau)
    \right]_{q^0}
    \in\bQ/\bZ,
\end{align}
where $[\cdot]_{q^{0}}$ denotes the coefficient of $q^{0}$.

The definition involves choices of the representatives $\sX$ and $\sT$, of $\sN$, of $(k,\sY_{0})$, and of $\sL$.
Under the conjectures above, it is shown in~\cite{Tachikawa:2025flw} that $W(\sX,\sT)\in\bQ/\bZ$ is independent of all of these choices and that $W$ is bilinear.
In our examples, we construct $\sN$, $(k,\sY_{0})$ and $\sL$ explicitly in Sections~\ref{sec:disk-bundle} and~\ref{sec:bordism}.

\subsection{Internal theories with a symmetry and their coupling to sigma models}
\label{sec:review-coupling}
In our examples, $\sX$ is the $\cN=(0,1)$ sigma model whose target space is a closed $d$-dimensional spin manifold $X$, so that $\nu_\sX=d$.
To define this theory, we need a string structure on $X$ to cancel the anomaly of the right-moving Majorana--Weyl fermions~\cite{Witten:1999eg}.
A string structure on $X$ is a trivialization of the characteristic class $\lambda(TX)=p_{1}(TX)/2$ introduced in~\eqref{eq:inflow-lambda}.
In the terminology of string theory, it is provided by the $B$-field of the sigma model.
At the level of differential forms, the field strength $H$ of the $B$-field satisfies $\d H=-\lambda(TX)$.
The string structure contains more information than the three-form $H$: two string structures on $X$ differ by a class in $H^3(X;\bZ)$, and only its image in de Rham cohomology is seen as the difference of their $H$.

The theory $\sT$ is an internal theory, an $\cN=(0,1)$ theory.
In our examples, it is a purely left-moving fermionic conformal field theory, regarded as an $\cN=(0,1)$ theory on which the supercharge acts trivially.
To define the tertiary invariant, $\nu_{\sT}=-22-d$ is required.
In our examples, $\sT$ has a symmetry $G$ realized by level-one current algebras.
Its 't~Hooft anomaly is characterized by a degree-four characteristic class $c$ of principal $G$-bundles.
In our examples, it is essentially sufficient to fix the conventions of $c$ for $G=\SU(r)$ and $G=\Spin(r)$.
We take connections to be anti-Hermitian, and extend the normalization \eqref{eq:inflow-level-one} to $G=\mathrm{SU}(r)$ and $\mathrm{Spin}(r)$ as
\begin{align}
\label{eq:review-8}
c=
\begin{cases}
\operatorname{ch}_2=\dfrac12\operatorname{tr}\Bigl(\dfrac{\i F}{2\pi}\Bigr)^2=-c_2,&G=\mathrm{SU}(r),\\[8pt]
\dfrac{1}{2}s_{1},&G=\mathrm{Spin}(r),
\end{cases}
\end{align}
where $\operatorname{tr}$ and $c_2$ refer to the fundamental representation of $\mathrm{SU}(r)$, and $s_j$ are the Pontryagin forms computed in the vector representation of $\mathrm{Spin}(r)$, as in Section~\ref{sec:inflow-ten-dimensional}.
For a product of groups, $c$ is the sum of the classes of its factors.
The sign is fixed as follows.
For $\widehat{\so}(r)_1$ realized by left-moving Majorana--Weyl fermions in the vector representation, $c=s_1/2$ takes the same form as $\lambda=p_1/2$ for the tangent bundle, so that the anomalies of the left- and right-moving fermions cancel when the gauge bundle in the vector representation is identified with the tangent bundle.
The case of $\widehat{\su}(r)_1$ follows by restriction to $\mathrm{SU}(r)\subset\mathrm{Spin}(2r)$, under which the vector representation becomes the fundamental representation regarded as real, and $s_1/2=-c_2$.

To compute the tertiary invariant, we need a noncompact theory $\sN$ with $\partial\sN=\sX\times\sT$.
To this end, we take a $(d+1)$-dimensional manifold $Y$ with $\partial Y=X$ and a principal $G$-bundle $P$ on $Y$ with a connection $A$.
We then couple $\sT$ to the sigma model on $Y$: we pull back $P$ and $A$ to the worldsheet along the sigma model field $\phi$, and couple $\phi^{\ast}A$ to the currents of the $G$ symmetry of $\sT$.
The anomaly of $\sT$ now adds to that of the sigma model, and the $B$-field must trivialize $\lambda(TY)-c$ instead of $\lambda(TY)$.
Such a trivialization is called a twisted string structure, and its field strength satisfies the Bianchi identity \eqref{eq:inflow-bulk-bianchi} on $Y$,
\begin{align}
\label{eq:review-9}
\d H=c-\lambda(TY).
\end{align}
We require that $P|_{X}$ is trivialized and that $A$ is pure gauge on $X$.
Then, $c$ vanishes on $X$, and the twisted string structure restricts to a string structure on $X$.
We require it to coincide with the string structure of the theory $\sX$, which itself need not extend over $Y$.
Since $\sT$ decouples near $X$, the resulting theory $\sN$ satisfies
\begin{align}
\label{eq:review-bdry}
    \partial\sN=\sX\times\sT.
\end{align}

The basic example is the three-sphere.
We equip the four-dimensional disk $D^4$ with a spherically symmetric metric, so that $\lambda(TD^4)=0$, and $S^3=\partial D^4$ with the induced round metric.
With this choice, $\int_{S^{3}}H$ is an integer for any string structure on $S^{3}$, and it vanishes exactly for the string structure that extends over $D^4$.
We consider the string structure with $\int_{S^{3}}H=1$, where the sign fixes the orientation of $S^{3}$.
It does not extend over $D^{4}$, but it extends as a twisted string structure.
Take a $G$-bundle on $D^4$ with a connection that is trivial near $S^3$, and orient $D^4$ as explained around \eqref{eq:inflow-stokes}.
Then, Stokes' theorem applied to \eqref{eq:review-9} gives
\begin{align}
\label{eq:review-10}
\int_{D^4}c=\int_{S^3}H=1.
\end{align}
This is the smooth counterpart of the flux \eqref{eq:inflow-flux}, with the source $\delta_\Sigma$ replaced by a gauge field.
We call $\int_{D^4}c$ the instanton number.
By \eqref{eq:inflow-level-one}, instanton number one is realized by a connection on an $\mathrm{SU}(2)$ subgroup of $G$ associated with a root.
In Section~\ref{sec:disk-bundle}, we realize such a connection by the BPST instanton, and apply this construction to every fiber of a sphere bundle.

\section{A disk bundle with an instanton in each fiber}
\label{sec:disk-bundle}

In this section, we construct a manifold $Y$ with $\partial Y=X$ by replacing each $S^3$ fiber of $X$ with a four-disk carrying an instanton, i.e.~by applying the basic example \eqref{eq:review-10} of Section~\ref{sec:review-coupling} to every fiber.
This raises two questions.
The first is how to put an instanton in every fiber consistently with changes of fiber coordinates.
We answer it over a general base in Section~\ref{sec:disk-bundle-general-base}.
The second is how to make the gauge field trivial on the boundary, so that the boundary theory is the product $\sX\times\sT$.
We answer it in Section~\ref{sec:disk-bundle-sphere-base} for the sphere bundles used in our examples.
Throughout, vertical directions are along the four-dimensional fiber, and horizontal directions are along the base.

\subsection{Instantons on a disk bundle over a general base}
\label{sec:disk-bundle-general-base}

Let $B$ be a smooth manifold without boundary, and let $V\to B$ be an oriented real four-plane bundle with a chosen spin lift of its rotation group.
Taking the unit sphere and the unit disk in each four-plane, we obtain the sphere and disk bundles\footnote{Since $X$ bounds the spin manifold $Y$, the ordinary and mod-2 elliptic genera of the sigma model on $X$ vanish. This does not assert that the string structure extends without the instanton; that obstruction is what the tertiary invariant detects.}
\begin{equation}
\label{eq:disk-1}
\begin{aligned}
S^3&\longrightarrow X=S(V)\longrightarrow B,\\
D^4&\longrightarrow Y=D(V)\longrightarrow B,
\end{aligned}
\qquad \partial Y=X.
\end{equation}
We want unit $H$-flux through each oriented sphere fiber, and the instanton on the disk bounded by that sphere will supply it.

The rotations of the fiber have the spin cover $\SU(2)_L\times\SU(2)_R$, and hence the spin lift specifies a principal $\SU(2)_L$ bundle and a principal $\SU(2)_R$ bundle on $B$.
These describe the geometry of $V$; they are not gauge bundles.
The actions of $\SU(2)_{L}$ and $\SU(2)_{R}$ become explicit when we write a fiber coordinate as a quaternion, represented by a matrix $y$ with $y^\dagger y=yy^\dagger=r^2\mathbf{1}_2$, where $r$ is the radial coordinate.
A change of local fiber frame then acts as
\begin{equation}\label{eq:disk-rotation}
y\longmapsto g_Lyg_R^{-1},\qquad
(g_L,g_R)\in\SU(2)_L\times\SU(2)_R.
\end{equation}
To write down the instanton, we use an unbounded radial coordinate on the interior of each disk and identify the interior with $\bH\simeq\bR^4$; the sphere at $r\to\infty$ represents the boundary of the disk.

We choose a gauge subgroup $\SU(2)_1$ of the symmetry group $G$ of $\sT$, with the level-one normalization of \eqref{eq:review-8}.
The subscript $1$ labels the gauge group of the instanton, while $L$ and $R$ label the rotations of the fiber.
On a single fiber, we take the BPST instanton~\cite{Belavin:1975fg} of size $\rho>0$,
\begin{equation}
\label{eq:disk-2}
A_{\rm inst}
=\frac{\tfrac12(y\,\d y^\dagger-\d y\,y^\dagger)}{r^2+\rho^2},
\qquad
F_{\rm inst}=\frac{\rho^2\,\d y\,\d y^\dagger}{(r^2+\rho^2)^2}.
\end{equation}
Its asymptotic value is $u\,\d u^{-1}$ with $u=y/r\in\SU(2)$.
Since $u$ has degree one as a map $S^3\to\SU(2)\simeq S^3$, the instanton number is one, and in our orientation the Bianchi identity on the fiber gives $\int_{S^3}H=\int_{D^4}c=1$, as in \eqref{eq:review-10}.
We use the BPST profile because its zero modes are explicitly known.
The construction only needs a smooth connection with instanton number one and this behavior at the boundary; in particular, it need not solve the Yang--Mills equation on $Y$.

Placing \eqref{eq:disk-2} independently in each fiber does not yet give a connection on $Y$.
On an overlap of patches of $B$, the transition functions $g_L$ and $g_R$ depend on the base point, so differentiating $g_Lyg_R^{-1}$ produces their derivatives.
These horizontal terms must be compensated so that the local instantons form a single gauge field.
For this purpose, let $A_L$ and $A_R$ be connections on the $\SU(2)_L$ and $\SU(2)_R$ bundles, pulled back from $B$ to $Y$, and introduce the covariant differential
\begin{equation}
Dy=\d y+A_Ly-yA_R.
\end{equation}
Under a change of frame, these quantities transform as
\begin{equation}
\begin{aligned}
y&\longmapsto g_Lyg_R^{-1},\\
A_a&\longmapsto g_aA_ag_a^{-1}+g_a\,\d g_a^{-1},
\qquad a=L,R,\\
Dy&\longmapsto g_L(Dy)g_R^{-1}.
\end{aligned}
\end{equation}
Therefore, $y(Dy)^\dagger-(Dy)y^\dagger$ transforms by conjugation with $g_L$, and adding a multiple of it to $A_L$ gives a connection with the same transformation law as $A_L$.
We fix the radial coefficient so that the instanton is recovered on each fiber:
\begin{equation}
\label{eq:disk-3}
\begin{aligned}
A_1&=A_L+
\frac{y(Dy)^\dagger-(Dy)y^\dagger}{2(r^2+\rho^2)}\\
&=\frac{\tfrac12(y\,\d y^\dagger-\d y\,y^\dagger)
+yA_Ry^\dagger+\rho^2A_L}{r^2+\rho^2},
\end{aligned}
\end{equation}
where the second line uses the anti-Hermiticity of $A_L$ and $A_R$ and $yy^\dagger=r^2\mathbf{1}_2$.
On a fiber, the pulled-back one-forms $A_L$ and $A_R$ vanish, and \eqref{eq:disk-3} reduces to \eqref{eq:disk-2}.
On overlaps, $A_1$ transforms as
\begin{equation}
A_1\longmapsto g_LA_1g_L^{-1}+g_L\,\d g_L^{-1}.
\end{equation}
Thus, $A_1$ is a connection on the $\SU(2)_1$ gauge bundle obtained by pulling back the $\SU(2)_L$ bundle through $\SU(2)_L\simeq\SU(2)_1$.

This gauge bundle is identified with different bundles near the core and near the boundary.
At the zero section $y=0$, $A_1$ reduces to $A_L$, and the instanton preserves only the diagonal subgroup $\SU(2)_D\subset\SU(2)_1\times\SU(2)_L$.
In the limit $r\to\infty$, on the other hand, the $A_L$ term is suppressed, and \eqref{eq:disk-3} becomes
\begin{equation}
\label{eq:disk-4}
A_1\longrightarrow u\,\d u^{-1}+uA_Ru^{-1}
\ \xrightarrow{\text{gauge transformation by }u^{-1}}\ A_R.
\end{equation}
The gauge-transformed connection $u^{-1}A_1u+u^{-1}\d u$, which approaches $A_R$, has transition functions $g_R$ instead of $g_L$.
Thus, near the boundary sphere, the $\SU(2)_1$ gauge bundle is also the pullback of the $\SU(2)_R$ bundle through $\SU(2)_R\simeq\SU(2)_1$.
These two identifications do not contradict each other.
On the total space $X$, the unit fiber coordinate $u$, which transforms as $u\mapsto g_Lug_R^{-1}$, identifies the pullbacks of the $\SU(2)_L$ and $\SU(2)_R$ bundles.
However, since $u$ is a function on $X$, not on $B$, the two bundles need not be isomorphic on $B$; for example, the $\SU(2)_L$ bundle can be trivial while the $\SU(2)_R$ bundle is not.

The difference between the core and the boundary is most transparent in the limit of a small instanton.
Consider the level-one four-form $c$ of \eqref{eq:inflow-level-one} computed from \eqref{eq:disk-3}.
Since only the BPST curvature has four components along the fiber, the integral of $c$ over each fiber disk is the instanton number one, independently of $A_L$ and $A_R$.
As $\rho\to0$, this fiber density is concentrated at the zero section and becomes a Thom form $\delta$ of $V$ as in \eqref{eq:inflow-source}.
Away from the core, at $r\gg\rho$, the connection is gauge equivalent to $A_R$ by \eqref{eq:disk-4}, and hence $c$ there is the pullback of $c(R)$, while the terms involving $A_L$ are suppressed by powers of $\rho^2/r^2$.
Therefore, we obtain
\begin{equation}
\label{eq:disk-c-limit}
c\longrightarrow\delta+c(R)\qquad(\rho\to0).
\end{equation}
In this limit, the instanton becomes an NS-brane on the zero section.
Since $A_1=A_L$ at $y=0$ for every $\rho$, the restriction of $c$ to $y=0$ is $c(L)$.
Restricting \eqref{eq:disk-c-limit} to $y=0$ therefore gives $\delta|_{y=0}=c(L)-c(R)=\chi(V)$ by \eqref{eq:inflow-normal-classes}, which is the self-intersection \eqref{eq:inflow-delta-euler}.

This limit also determines the gauge field to which the theory on the brane couples.
Comparing \eqref{eq:disk-c-limit} with \eqref{eq:inflow-bianchi-source}, the brane theory couples to the gauge field at $r\to\infty$, which is $A_R$ up to gauge, rather than to its value $A_L$ at the core.
For the supersymmetric strings, this is consistent with the Higgs branch of the brane theory, which describes instantons of finite size: the expectation values of the hypermultiplet scalars, which carry $\SU(2)_R$ charge, break $\SU(2)_1\times\SU(2)_R$ to its diagonal subgroup.\footnote{Which diagonal subgroup is preserved is a confusing point. In the gauge \eqref{eq:disk-2}, which is regular at the core, a left rotation $y\mapsto g_Ly$ is compensated by the gauge transformation $g_L$, giving $\SU(2)_D\subset\SU(2)_1\times\SU(2)_L$, while right rotations leave the field invariant. After the gauge transformation by $u^{-1}$ in \eqref{eq:disk-4}, which is regular at fiber infinity, the roles are exchanged: left rotations leave the field invariant, and a right rotation $y\mapsto yg_R^{-1}$ is compensated by the gauge transformation $g_R$, giving the diagonal subgroup of $\SU(2)_1\times\SU(2)_R$. The two descriptions differ by the position-dependent gauge transformation $u$, so there is no contradiction; the flavor symmetry of the brane theory refers to the frame at fiber infinity.}

We now return to the construction.
The boundary condition required in Section~\ref{sec:review-pairing} is that the boundary theory is the product $\sX\times\sT$, as in \eqref{eq:review-bdry}.
Hence, the $G$-connection on $X$ must be made gauge equivalent to zero without losing the unit instanton number on each fiber.
In the frame of \eqref{eq:disk-4}, the term $u\,\d u^{-1}$ carrying the instanton number has been gauged away, and the boundary connection is $A_R$ pulled back from $B$, which is in general a nonzero background for $\sT$.
Let $P_G$ be the principal $G$-bundle on $B$ obtained from the $\SU(2)_R$ bundle through $\SU(2)_R\simeq\SU(2)_1\subset G$, with the connection induced from $A_R$.
If $P_G$ is trivial, this connection can be deformed to zero on $B$ in a trivialization, and this suffices to remove the background.
Since only $P_G$ needs to be trivial, not the $\SU(2)_R$ bundle itself, $V$ and $X$ can remain nontrivial.
We now implement this for a sphere base.

\subsection{Trivialization of the gauge field on the boundary}
\label{sec:disk-bundle-sphere-base}

In this subsection, we trivialize the gauge field on the boundary when the base is a sphere.
From now on, we impose the following assumptions:
\begin{enumerate}

\item $B=S^n$ with $n>4$, so that $d=\dim X=n+3$.

\item The $\SU(2)_L$ bundle is trivial, and we choose its connection to be $A_L=0$.

\item The $\SU(2)_R$ bundle on $B$ becomes trivial when its structure group is embedded in $G$ through $\SU(2)_R\simeq\SU(2)_1\subset G$.

\end{enumerate}
The condition $n>4$ will be used in Section~\ref{sec:bordism-commutant}.
By the second assumption, the $\SU(2)_1$ gauge bundle, which is the pullback of the $\SU(2)_L$ bundle, is trivial on $Y$, and \eqref{eq:disk-3} is a global one-form in this trivialization.
The third assumption says that the bundle $P_G$ of Section~\ref{sec:disk-bundle-general-base} is trivial.

The third assumption requires triviality on $B$, although the product $\sX\times\sT$ might seem to require only a trivialization on $X$.
One might deform the boundary value of \eqref{eq:disk-3} to zero in the trivialization given by the second assumption, which would give the product boundary condition.
However, the instanton number on each fiber, measured relative to this boundary trivialization, would then be zero, and so would the flux $\int_{S^3}H$.
To keep the unit instanton number, the boundary field must instead be trivialized in the frame of \eqref{eq:disk-4}, up to a gauge transformation pulled back from $B$.
In that frame, the remaining boundary connection is $A_R$ pulled back from $B$, so such a trivialization exists when $P_G$ is trivial on $B$.

We write $B=S^n=D^n_{\north}\cup D^n_{\south}$, where $D^n_{\north}$ and $D^n_{\south}$ are the northern and southern hemispheres, which meet along the equator $S^{n-1}$.
We trivialize the $\SU(2)_R$ bundle over $D^n_{\north}$ and $D^n_{\south}$.
The topology of the $\SU(2)_{R}$ bundle is then encoded in the transition function on the equator, called the clutching function:
\begin{equation}
h:S^{n-1}\longrightarrow\SU(2)_R,
\qquad y_{\south}=y_{\north}h,
\end{equation}
where $y_{\north}$ and $y_{\south}$ are the fiber coordinates $y$ of $V$ in the northern and southern trivializations.
The homotopy class $[h]\in\pi_{n-1}(\SU(2)_R)$ determines the $\SU(2)_R$ bundle on $S^n$ up to isomorphism.
Requiring $Dy=\d y+A_Ly-yA_R$ to transform in the same way as $y$, we find that the $\SU(2)_R$ connection obeys
\begin{equation}
A_R^{\south}
=h^{-1}A_R^{\north}h+h^{-1}\d h.
\end{equation}

We now construct a global trivialization of $P_G$ from the third assumption.
We denote the connection on $P_G$ by $A_G$.
Initially, $A_G$ is the image of $A_R$; we will deform it into a $G$-connection that does not come from any $\SU(2)_R$ connection, while keeping the $\SU(2)_R$ bundle, and hence $V$ and $X$, fixed.
In the trivializations of $P_G$ induced from the northern and southern trivializations of the $\SU(2)_R$ bundle, the transition function of $P_G$ is $h$, now regarded as $\SU(2)_1$-valued, and $A_G^{\south}=h^{-1}A_G^{\north}h+h^{-1}\d h$.
By the third assumption, $[h]$ maps to zero in $\pi_{n-1}(G)$.
Therefore, $h$ can be contracted in $G$, i.e.~extended from the equator to the northern hemisphere:
\begin{equation}
\label{eq:disk-5}
h_1:D^n_{\north}\longrightarrow G,\qquad
h_1|_{S^{n-1}}=h.
\end{equation}
We change the northern frame by the gauge transformation $h_1$, which is defined on all of $D^n_{\north}$, and keep the southern frame.
In the new frames, $A_G$ is expressed as
\begin{equation}
A'_G=\begin{cases}
h_1^{-1}A_G^{\north}h_1+h_1^{-1}\d h_1,
&\text{on }D^n_{\north},\\
A_G^{\south},&\text{on }D^n_{\south}.
\end{cases}
\end{equation}
Since $h_1=h$ on the equator, the new northern frame and the southern frame agree there.
Thus, the new frames form a single trivialization of $P_G$ over $S^n$, in which $A'_G$ is a globally defined $\g$-valued one-form; it is still the same connection as $A_G$, i.e.~the image of $A_R$.

This trivialization allows us to deform the connection to zero.
Since $A'_G$ is a global one-form, $(1-s)A'_G$ with $s\in[0,1]$ is a family of connections on $P_G$.
In the original frames, it gives the homotopy of connections
\begin{equation}
\label{eq:disk-homotopy}
A_{G,s}=\begin{cases}
(1-s)A_R^{\north}+s\,h_1\,\d h_1^{-1},
&\text{on }D^n_{\north},\\
(1-s)A_R^{\south},&\text{on }D^n_{\south},
\end{cases}
\end{equation}
which satisfies $A_{G,s}^{\south}=h^{-1}A_{G,s}^{\north}h+h^{-1}\d h$ for every $s$.
It connects $A_{G,0}$, the image of $A_R$, to the trivial connection $A_{G,1}$ with $A_{G,1}^{\north}=h_1\,\d h_1^{-1}$ and $A_{G,1}^{\south}=0$.

Finally, we interpolate along the radial direction $r$ of the fiber between the instanton near the core and the trivial connection near the boundary, keeping $P_G$ and its transition function $h$ fixed.
With $A_L=0$ and $y=ru$, the instanton field \eqref{eq:disk-3} is
\begin{align}
\label{eq:disk-A1-horizontal}
A_1=\frac{r^2}{r^2+\rho^2}\left(u\,\d u^{-1}+uA_Ru^{-1}\right)
=u\,\d u^{-1}+uA_1^\infty u^{-1},
\end{align}
where
\begin{equation}
\label{eq:disk-A1u}
A_1^\infty =\frac{r^2A_R+\rho^2\,u^{-1}\d u}{r^2+\rho^2}.
\end{equation}
Here, $A_1^\infty$ is the gauge field in the frame of \eqref{eq:disk-4}.
We choose a function $s(r)$ with $s=0$ near $r=0$ and $s=1$ near fiber infinity, and define a new gauge field by
\begin{align}
\label{eq:disk-deformed}
A&=u\,\d u^{-1}+uA^\infty u^{-1},\\
\label{eq:disk-deformed-frame-R}
A^\infty&=\frac{r^2A_{G,s(r)}+\rho^2\,u^{-1}\d u}{r^2+\rho^2}.
\end{align}
Near $r=0$, we have $A^\infty=A_1^\infty$ and $A=A_1$, and hence the instanton core is unchanged.
As $r\to\infty$, we have $A^\infty\to A_{G,1}$, and $A$ becomes pure gauge, $A|_X=t\,\d t^{-1}$, where $t=u_{\north}h_1$ on $D^n_{\north}$ and $t=u_{\south}$ on $D^n_{\south}$.
The two expressions of $t$ agree on the equator since $u_{\south}=u_{\north}h$.
Thus, $t$ is the global trivialization of $P_G$ given by $h_1$, composed with the gauge transformation $u$ of \eqref{eq:disk-4}.
Since $A_{G,s(r)}$ has only horizontal components, the restriction of $A$ to each fiber is still the instanton \eqref{eq:disk-2}, and on each fiber $t$ is $u$ times a constant element of $G$.
Therefore, the instanton number is one.

This completes the construction of the gauge field on $Y$, together with its trivialization on $X$.
Coupling the sigma model on $Y$ to $\sT$ by this field, as in Section~\ref{sec:review-coupling}, gives the noncompact theory $\sN$, whose boundary is $\sX\times\sT$ by \eqref{eq:review-bdry}.

\section{The construction of the target space}
\label{sec:bordism}

The theory $\sN$ constructed in Section~\ref{sec:disk-bundle} is the first of the three theories in the definition of the tertiary invariant in Section~\ref{sec:review-pairing}.
The other two are a theory $\sY_0$ with $\partial\sY_0=\sX^{\sqcup k}$ and a theory $\sL$ with $\partial\sL=\sN^{\sqcup k}\cup\overline{\sY_0\times\sT}$, as in \eqref{eq:review-2}.
In this section, we construct these two theories as sigma models with target spaces $Y_0$ and $Z$ satisfying $\partial Y_0=X^{\sqcup k}$ and $\partial Z=Y^{\sqcup k}\cup\overline{Y_0}$; the internal theory $\sT$ is coupled to the latter but not to the former.

\subsection{The construction of \texorpdfstring{$Y_0$}{Y0}}
\label{sec:bordism-joining}

We construct $Y_0$ with $\partial Y_0=X^{\sqcup k}$ by extending the base $B=S^{n}$ of $X$ to a manifold with boundary, keeping the $S^3$ fibers.
Since $\sT$ is not coupled to the sigma model on $Y_0$, no gauge field is needed there.
However, the $S^3$ fibers are the unit spheres of $V$, so the $\SU(2)_R$ bundle must extend over the new base within its own structure group $\SU(2)_R$, not merely after the embedding into $G$.
This is in general impossible for a single copy of $B$, but becomes possible for $k$ copies of $B$.

We assume that $[h]$ is a torsion element of $\pi_{n-1}(\SU(2)_R)$, and take $k>0$ such that $k[h]=0$; we will have $k=12$ in Section~\ref{sec:sp2} and $k=2$ in Section~\ref{sec:su3}.
The construction below realizes the sum $k[h]$ geometrically, by joining the $k$ base spheres together with their bundles.

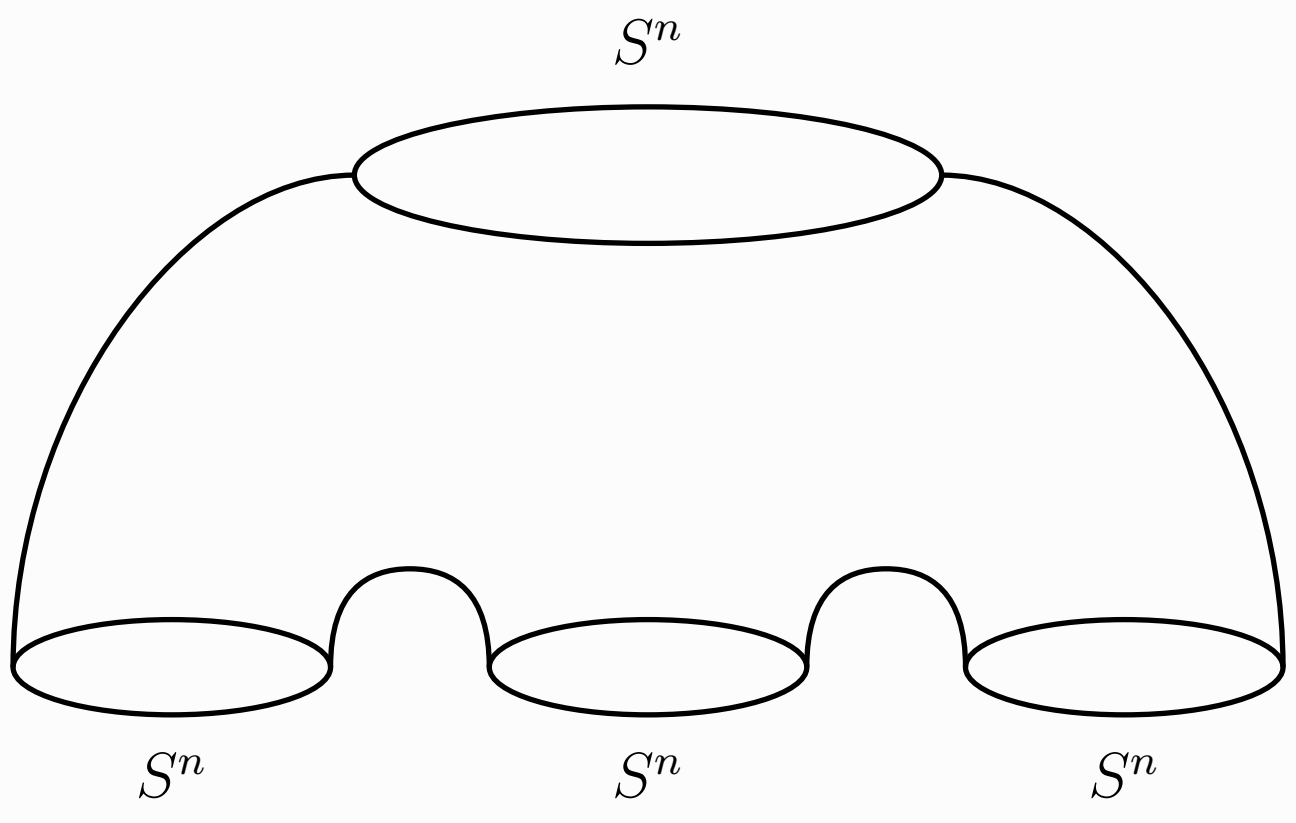
\begin{figure}[htbp]
\centering
\begin{tikzpicture}[x=1.05cm,y=1.05cm,line width=1pt]
  \draw (-1,0) .. controls (-1,1.65) and (0.10,3.1) .. (1.15,3.1);
  \draw (7,0) .. controls (7,1.65) and (5.90,3.1) .. (4.85,3.1);
  \draw (1,0) .. controls (1,.47) and (1.22,.62) .. (1.5,.62)
               .. controls (1.78,.62) and (2,.47) .. (2,0);
  \draw (4,0) .. controls (4,.47) and (4.22,.62) .. (4.5,.62)
               .. controls (4.78,.62) and (5,.47) .. (5,0);
  \draw (0,0) ellipse (1 and .30);
  \draw (3,0) ellipse (1 and .30);
  \draw (6,0) ellipse (1 and .30);
  \draw (3,3.1) ellipse (1.85 and .43);
  \node at (0,-.68) {$S^n$};
  \node at (3,-.68) {$S^n$};
  \node at (6,-.68) {$S^n$};
  \node at (3,3.94) {$S^n$};
\end{tikzpicture}
\space 
\caption{The joining part of the base bordism, illustrated for $k=3$. The three lower openings represent the original $S^n$ components; the upper opening represents their connected sum. Attaching a disk to the upper opening, in the next step, leaves just the three original boundary components. The drawing represents the base geometry; the $S^3$ fibers are not shown.}
\label{fig:joining}
\end{figure}

We first build a bordism from the disjoint union of $k$ spheres $S^n$ to their connected sum, by joining them with thin handles attached at points on their equators:
\begin{equation}
\bigsqcup_{i=1}^kS^n
\xrightarrow{\text{join by handles}}
S^n\#\cdots\#S^n\simeq S^n.
\end{equation}
This bordism is shown in Figure~\ref{fig:joining}.

Since the handles are attached at points on the equators, the $k$ northern hemispheres $D^n_{\north}$ join into a northern hemisphere of the resulting sphere, which we again denote by $D^n_{\north}$, and similarly for $D^n_{\south}$.
The equators form the connected sum $S^{n-1}\#\cdots\#S^{n-1}\simeq S^{n-1}$, which is the equator of the resulting sphere, and the $\SU(2)_R$ bundles of the $k$ copies join into a bundle on the resulting sphere.
We denote its clutching function on this equator by $h^{\#k}$.
Under the connected sum, the classes of clutching functions add in $\pi_{n-1}(\SU(2)_R)$, and hence $[h^{\#k}]=k[h]=0$.
Therefore, the $\SU(2)_R$ bundle on the resulting sphere is trivial, and it extends over a disk $D^{n+1}$ attached to that sphere.
We call the attached disk the cap, and denote by $C$ the union of the bordism and the cap.
It is diffeomorphic to $S^{n+1}$ with $k$ balls removed, and
\begin{equation}
  \partial C=(S^n)^{\sqcup k}.
\end{equation}

\begin{figure}[htbp]
\centering
\begin{tikzpicture}[x=1cm,y=1cm,line width=1pt,font=\small]
  \fill[SteelBlue!10] (2.2,0) arc[start angle=0,end angle=180,radius=2.2] -- cycle;
  \fill[gray!8] (-2.2,0) arc[start angle=180,end angle=360,radius=2.2] -- cycle;
  \draw[SteelBlue,line width=1.3pt] (2.2,0) arc[start angle=0,end angle=180,radius=2.2];
  \draw[SlateGray,line width=1.3pt] (-2.2,0) arc[start angle=180,end angle=360,radius=2.2];
  \draw[Sienna,line width=1.2pt] (-2.2,0) -- (2.2,0);
  \fill (-2.2,0) circle (1.8pt);
  \fill (2.2,0) circle (1.8pt);
  \node at (0,1.13) {$D^{n+1}_{\north}$};
  \node at (0,-1.13) {$D^{n+1}_{\south}$};
  \draw[SteelBlue,dashed] (0,1.72) circle (0.32);
  \draw[SteelBlue,->,shorten >=3pt] (2.35,2.25) -- (0.3,1.83);
  \node[SteelBlue,anchor=south west,align=left] at (2.25,2.2) {$D^{n+1}_{\rm in}$\\$S^n_{\rm in}:g'$};
  \node[Sienna,above=1pt,inner sep=0pt] at (0,0) {$D^n_{\rm equator}:h_2$};
  \draw[SteelBlue,->,shorten >=3pt] (-3.25,2.14) -- (-1.58,1.54);
  \node[SteelBlue,anchor=south east] at (-3.15,2.14) {$D^n_{\north}:h_1^{\# k}$};
  \draw[SlateGray,->,shorten >=3pt] (-3.28,-2.10) -- (-1.58,-1.54);
  \node[SlateGray,anchor=north east] at (-3.18,-2.10) {$D^n_{\south}$};
  \draw[->,shorten >=5pt] (3.05,-.04) -- (2.2,0);
  \node[anchor=west] at (3.05,-.04) {$S^{n-1}:h^{\#k}$};
\end{tikzpicture}
\space 
\caption{A cross-section of the cap $D^{n+1}$ attached to the connected sum in Figure~\ref{fig:joining}. The upper and lower outer arcs represent the hemispheres $D^n_{\north}$ and $D^n_{\south}$ of the connected sum, and the interior diameter represents $D^n_{\rm equator}$. They meet along the equator $S^{n-1}$, where the transition function is $h^{\#k}$; the map $h_2$ extends it over $D^n_{\rm equator}$. The upper arc and the diameter together form the boundary $S^n$ of the northern half-ball $D^{n+1}_{\north}$, on which the map $g$ of \eqref{eq:bordism-2} is defined. The dashed disk is the ball $D^{n+1}_{\rm in}$ of Section~\ref{sec:bordism-commutant}; on its boundary $S^n_{\rm in}$, $g$ has been deformed to $g'$.}
\label{fig:cap}
\end{figure}
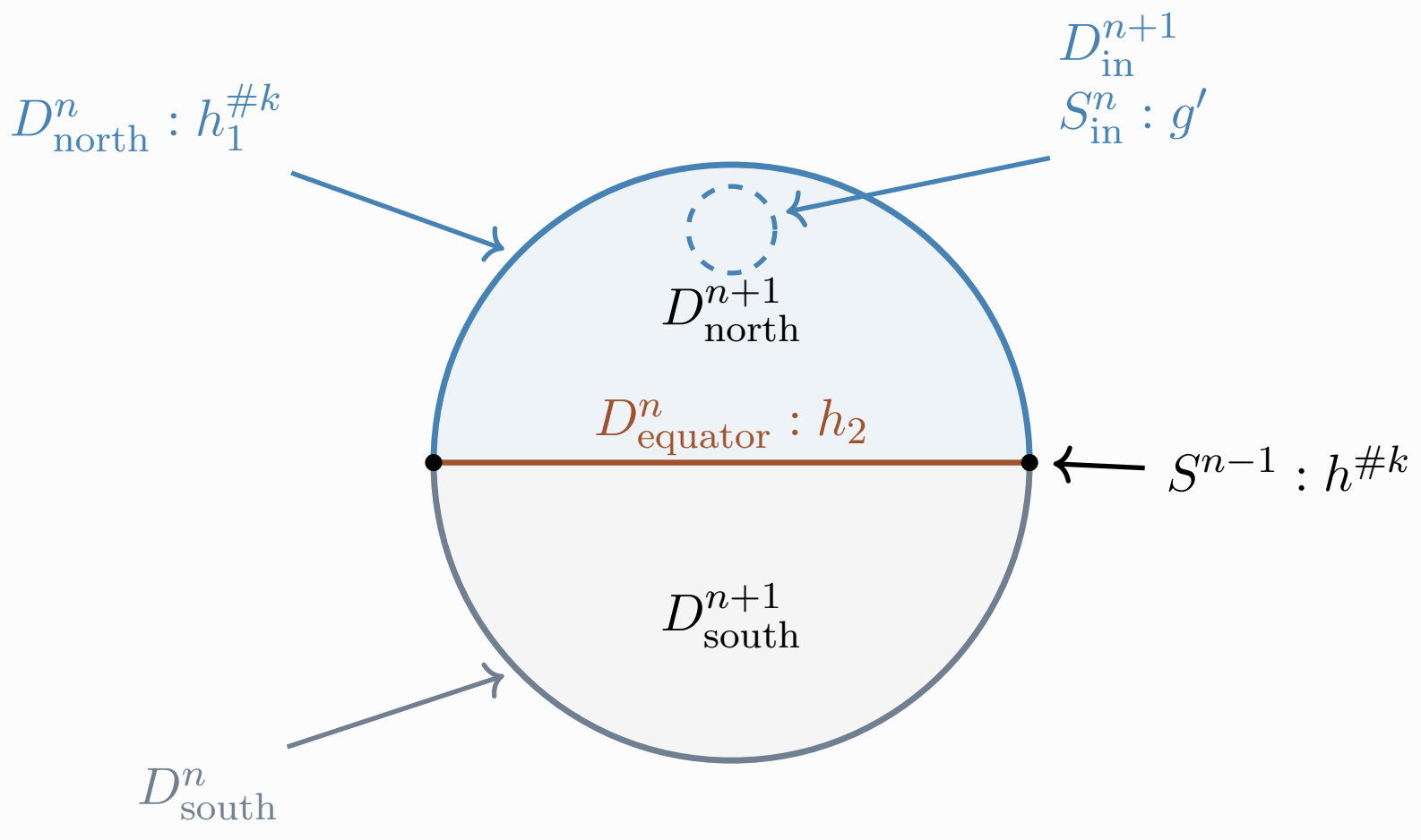

To describe the extension explicitly, we divide the cap into the northern and southern half-balls $D^{n+1}_{\north}$ and $D^{n+1}_{\south}$, as in Figure~\ref{fig:cap}.
The half-balls meet along a disk $D^n_{\rm equator}$ bounded by the equator $S^{n-1}$ of the resulting sphere, so that $\partial D^{n+1}_{\north}=D^n_{\north}\cup D^n_{\rm equator}$ and $\partial D^{n+1}_{\south}=D^n_{\south}\cup D^n_{\rm equator}$.
Since $k[h]=0$, the clutching function $h^{\#k}$ extends to
\begin{equation}
\label{eq:bordism-1}
h_2:D^n_{\rm equator}\longrightarrow\SU(2)_R,\qquad
h_2|_{S^{n-1}}=h^{\#k}.
\end{equation}
We take a trivial four-plane bundle on each half-ball, and identify their fiber coordinates on $D^n_{\rm equator}$ by $y_{\south}=y_{\north}h_2$, which reduces to $y_{\south}=y_{\north}h^{\#k}$ on $S^{n-1}$.
Together with the $k$ copies of $V$ joined over the bordism, this defines a four-plane bundle $V_C\to C$ with $V_C|_{\partial C}=V^{\sqcup k}$.
Up to orientation, $Y_0$ is its sphere bundle $S(V_C)$.
The string structures on the $k$ copies of $X$ extend over $Y_0$, as we will show in Section~\ref{sec:bordism-commutant}.

We now fix the orientation of $Y_0$ so that $\partial Y_0=X^{\sqcup k}$, as required in Section~\ref{sec:review-pairing}.
We use the conventions of Section~\ref{sec:anomaly-inflow}.
A boundary is oriented by the outward-normal-first rule \eqref{eq:inflow-stokes}, under which
\begin{equation}
\label{eq:bordism-product-boundary}
\partial(M_1\times M_2)
=\partial M_1\times M_2\ \cup\ (-1)^{\dim M_1}\,M_1\times\partial M_2 .
\end{equation}
The total space of a sphere bundle is oriented fiber first, as in Section~\ref{sec:inflow-ns5}, with the fiber orientation of \eqref{eq:review-10}.
We orient $C$ so that $\partial C=+(S^n)^{\sqcup k}$, where each $S^n$ carries the orientation of the base $B$ used for $X$.
Since $S(V_C)$ is locally $S^3\times C$, we have
\begin{equation}
\partial S(V_C)=(-1)^{3}\,S(V_C)|_{\partial C}=-X^{\sqcup k},
\end{equation}
where the sign appears because the fiber $S^3$ has odd dimension.
We therefore define $Y_0$ to be $S(V_C)$ with the reversed orientation, $Y_0=\overline{S(V_C)}:=-S(V_C)$, which satisfies $\partial Y_0=X^{\sqcup k}$.

\subsection{The construction of \texorpdfstring{$Z$}{Z}}
\label{sec:bordism-z}

The manifold $Z$ combines the constructions of $Y$ and $Y_0$: it has disk fibers over $C$.
Its gauge field must agree with the one constructed in Section~\ref{sec:disk-bundle} on each copy of $Y$ in $\partial Z$, and must be pure gauge on $Y_0$.
We first construct the manifold, and then solve this extension problem for the gauge field.

\subsubsection{The manifold \texorpdfstring{$Z$}{Z}}

We take $Z$ to be the disk bundle $D(V_C)$ of the four-plane bundle $V_C\to C$ of Section~\ref{sec:bordism-joining}.
Table~\ref{tab:spaces} summarizes the four spaces in our construction.

\begin{table}[htbp]
\centering
\small
\renewcommand{\arraystretch}{1.22}
\setlength{\tabcolsep}{4pt}
\begin{tabular}{@{}p{.27\linewidth}p{.12\linewidth}p{.10\linewidth}p{.41\linewidth}@{}}
\hline
\raggedright Space & \raggedright Base & \raggedright Fiber & \raggedright Role \tabularnewline
\hline
\raggedright $X=S(V)$ & \raggedright $B=S^n$ & \raggedright $S^3$ & \raggedright Original boundary target \tabularnewline
\raggedright $Y=D(V)$ & \raggedright $B=S^n$ & \raggedright $D^4$ & \raggedright Bounds $X$, with an instanton in each fiber \tabularnewline
\raggedright $Y_0=\overline{S(V_C)}$ & \raggedright $C$ & \raggedright $S^3$ & \raggedright Bounds $k$ copies of $X$, with $\sT$ not coupled \tabularnewline
\raggedright $Z=D(V_C)$ & \raggedright $C$ & \raggedright $D^4$ & \raggedright Has boundary components $Y^{\sqcup k}$ and $\overline{Y_0}$ \tabularnewline
\hline
\end{tabular}
\caption{The four spaces: the target $X$, the manifolds $Y$ and $Y_0$ with boundaries $X$ and $X^{\sqcup k}$, and the manifold $Z$ with boundary $Y^{\sqcup k}\cup\overline{Y_0}$.}
\label{tab:spaces}
\end{table}

The boundary of $Z$ consists of two parts.
Over $\partial C$, the disk bundle restricts to $D(V_C)|_{\partial C}=Y^{\sqcup k}$.
Over $C$, the boundaries of the disk fibers form the sphere bundle $S(V_C)$, which is $Y_0$ up to orientation.
The two parts meet along the corner $S(V_C)|_{\partial C}=X^{\sqcup k}$.

We orient $Z=D(V_C)$ fiber first, as we did for $S(V_C)$ in Section~\ref{sec:bordism-joining}.
Since $Z$ is locally $D^4\times C$, \eqref{eq:bordism-product-boundary} gives
\begin{equation}
\partial Z=D(V_C)|_{\partial C}\ \cup\ S(V_C)
=Y^{\sqcup k}\cup S(V_C) = Y^{\sqcup k}\cup\overline{Y_0},
\end{equation}
where no sign appears because the fiber $D^4$ has even dimension.
This is the boundary required by \eqref{eq:review-2}.
Rounding the corner along $X^{\sqcup k}$ gives a manifold with boundary $Y^{\sqcup k}\cup_X\overline{Y_0}$ and does not change these orientations.

\subsubsection{Transition functions and the obstruction}
\label{sec:bordism-obstruction}

In this subsection, we show that the gauge bundle on $Z$ cannot in general be obtained by repeating the construction of Section~\ref{sec:disk-bundle-sphere-base} over $C$.
For this purpose, we describe the transition functions on the cap explicitly, in the notation of Section~\ref{sec:bordism-joining}.

Besides the extension $h_2$ of $h^{\#k}$ in \eqref{eq:bordism-1}, which describes $V_C$ over the cap, the trivialization of $P_G$ over the $k$ copies of $B$ involves another extension of $h^{\#k}$.
Recall that changing the northern frame of $P_G$ by the contraction $h_1:D^n_{\north}\to G$ of $h$ in \eqref{eq:disk-5} trivializes $P_G$ over $B$.
Under the joining of the $k$ spheres, the $k$ maps $h_1$ combine into $h_1^{\#k}:D^n_{\north}\to G$.
Both $h_1^{\#k}$ and $h_2$ restrict to $h^{\#k}$ on the equator $S^{n-1}$, but they play different roles.
The map $h_1^{\#k}$ is defined on the northern hemisphere $D^n_{\north}$ and takes values in $G$; it trivializes $P_G$ over the $k$ copies of $B$.
The map $h_2$ is defined on the interior disk $D^n_{\rm equator}$ and takes values in $\SU(2)_R$; it extends the $\SU(2)_R$ bundle across the cap.
Gluing these two maps along $S^{n-1}$ gives
\begin{equation}
\label{eq:bordism-2}
g:S^n=D^n_{\north}\cup D^n_{\rm equator}\longrightarrow G,
\end{equation}
equal to $h_1^{\#k}$ on $D^n_{\north}$ and to $h_2$, regarded as $\SU(2)_1$-valued, on $D^n_{\rm equator}$.
The map $g$ measures the difference between the two extensions.
Its domain is $\partial D^{n+1}_{\north}$, shown in Figure~\ref{fig:cap}.

We now describe the gauge bundle in terms of frames, starting with $Y$.
In each fiber $D^4$ of $Y$, we distinguish the region near the instanton core from the region away from it, which includes fiber infinity.
\begin{itemize}
\item Near the instanton core, the gauge bundle is the pullback of the $\SU(2)_L$ bundle through $\SU(2)_L\simeq\SU(2)_1$, as in \eqref{eq:disk-3}. Since the $\SU(2)_L$ bundle is trivial, a single frame covers this region, and the gauge field \eqref{eq:disk-deformed} is written in it.
\item Away from the instanton core, the gauge bundle is the pullback of the $\SU(2)_R$ bundle, as in \eqref{eq:disk-4}, and its frame is related to the one near the core by the transition function $u$. In this frame, the connection \eqref{eq:disk-deformed} is $A^\infty$ of \eqref{eq:disk-deformed-frame-R}.
\item Away from the instanton core, there are, in addition, northern and southern frames over $D^n_{\north}$ and $D^n_{\south}$, related by the transition function $h$ on the equator.
\item Changing the northern frame by $h_1$ of \eqref{eq:disk-5} removes this transition function and gives a single frame of $P_G$ over all of $B$. In this frame, the homotopy \eqref{eq:disk-homotopy} deforms $A^\infty$ to zero at fiber infinity.
\end{itemize}
On $Y_0$, the gauge bundle is trivial by construction.
Therefore, the gauge bundle on $\partial Z=Y^{\sqcup k}\cup\overline{Y_0}$ has already been constructed.

Suppose that the gauge bundle on all of $Z$ could be taken to be the pullback of the $\SU(2)_L$ bundle near the instanton core and of the $\SU(2)_R$ bundle away from it, as on $Y$.
It suffices to consider the cap $D^{n+1}$.
There, the northern and southern frames away from the core, over $D^{n+1}_{\north}$ and $D^{n+1}_{\south}$, are related on $D^n_{\rm equator}$ by $h_2$, regarded as $\SU(2)_1$-valued.
A single frame would be obtained by changing the northern frame by a map $g_1:D^{n+1}_{\north}\to G$.
To remove the transition function, $g_1$ must equal $h_2$ on $D^n_{\rm equator}$.
To match the trivialization by $h_1$ on the copies of $X$, which is required for the gauge bundle to be trivial on $Y_0$, it must also equal $h_1^{\#k}$ on $D^n_{\north}$.
Therefore, $g_1$ restricts to $g$ of \eqref{eq:bordism-2} on $\partial D^{n+1}_{\north}$, and such $g_1$ exists precisely when $g$ extends over $D^{n+1}_{\north}$, i.e.~when $[g]=0\in\pi_{n}(G)$.

If $[g]\neq0$, no such $g_1$ exists.
We can still change the northern frame by $g$ on a collar of $\partial D^{n+1}_{\north}$ in $D^{n+1}_{\north}$, extending $g$ constantly in the inward direction, but not on all of $D^{n+1}_{\north}$.
Over $D^{n+1}_{\north}$, we then have two frames away from the instanton core: the frame changed by $g$ on the outer collar, which matches the rest of $C$, and the original northern frame on the inner part.
The map $g$ serves as the transition function between them.

The class $[g]$ is related to the original transition function by the homotopy exact sequence of $\SU(2)_1\to G\to G/\SU(2)_1$:
\begin{equation}
\label{eq:bordism-3}
\pi_n(G)\longrightarrow\pi_n(G/\SU(2)_1)
\xrightarrow{\partial}\pi_{n-1}(\SU(2)_1)
\longrightarrow\pi_{n-1}(G).
\end{equation}
Projecting $h_1$ to $G/\SU(2)_1$ maps its boundary to a point, since $h$ takes values in $\SU(2)_1$.
Collapsing the boundary, we obtain the relative class $[h_1]\in\pi_n(G/\SU(2)_1)$, and the boundary homomorphism gives $\partial[h_1]=[h]$.
For $k$ copies, $\partial(k[h_1])=k[h]=0$, and hence by exactness $k[h_1]$ is the image of a class in $\pi_n(G)$.
The map $g$ represents such a class; in this sense, $[g]$ is a lift of $k[h_1]$.
This lift depends on the choices of $h_1$ and $h_2$, and we compute it in each example in Sections~\ref{sec:sp2} and~\ref{sec:su3} by using the homomorphisms in \eqref{eq:bordism-3}.

Therefore, if $[g]\neq0$, the gauge bundle on $Z$ cannot be taken to be the pullback determined by $V_C$: that pullback restricts to the gauge bundle on $Y^{\sqcup k}$, but not to the trivial one on $Y_0$.
The gauge bundle on $Z$ must differ from it in the interior, and the construction of Section~\ref{sec:disk-bundle-sphere-base} cannot simply be repeated over $C$.
In Section~\ref{sec:bordism-commutant}, we construct such a gauge bundle, with a connection that may have curvature in the interior and is pure gauge at fiber infinity.

\subsubsection{The gauge bundle and connection}
\label{sec:bordism-commutant}

Let $G'$ be the \emph{commutant} of $\SU(2)_1$ in $G$, i.e.~the subgroup consisting of elements that commute with every element of $\SU(2)_1$.
A $G'$ gauge field can act on the instanton zero modes without changing the instanton background.
Suppose that the transition function $g$ of Section~\ref{sec:bordism-obstruction} can be replaced by a $G'$-valued map $g'$.
Then, we can move $g'$ from fiber infinity to the region near the instanton core without changing the instanton.
There, only the one-form $g'^{-1}\d g'$ needs to be extended over the interior of $D^{n+1}_{\north}$, and this is possible even when the map $g'$ itself does not extend.
The resulting connection has curvature near the core.

We therefore assume that $g$ can be continuously deformed in $G$ to a map $g':S^n\to G'$, i.e.~that $[g]$ is in the image of $\pi_n(G')\to\pi_n(G)$.
This is an additional condition, to be verified in each example; it does not follow from the existence of $h_1$ and $h_2$ alone.
We write the deformation as $g_\sigma$ with $\sigma\in[0,1]$, and take $g_\sigma=g$ for $\sigma$ near zero and $g_\sigma=g'$ for $\sigma$ near one.

We first construct the gauge field outside $D^{n+1}_{\north}$.
We extend the $\SU(2)_R$ connection $A_R$ on the $k$ copies of $B$ to a connection on the $\SU(2)_R$ bundle of $V_C$; any extension will do.
We also denote by $P_G$ the $G$-bundle on $C$ obtained from this $\SU(2)_R$ bundle through $\SU(2)_R\simeq\SU(2)_1\subset G$, and by $A_G$ its connection induced from $A_R$.
Outside $D^{n+1}_{\north}$, we change the northern frame by $h_1^{\#k}$, extended along the bordism direction, and keep the southern frame; as on $Y$, these frames form a single trivialization of $P_G$.
We define $A_{G,s}$ by \eqref{eq:disk-homotopy} with $h_1$ replaced by $h_1^{\#k}$, and the gauge field by \eqref{eq:disk-deformed} and \eqref{eq:disk-deformed-frame-R}.

We now turn to $D^{n+1}_{\north}$.
We choose a smaller $(n+1)$-ball $D^{n+1}_{\rm in}$ in the interior of $D^{n+1}_{\north}$ as in Figure~\ref{fig:cap}, with boundary $S^n_{\rm in}$, and identify $D^{n+1}_{\north}\setminus D^{n+1}_{\rm in}$ with $[0,1]\times S^n$, with $\sigma$ increasing inward.
On this collar, we change the northern frame away from the instanton core by $g_\sigma$.
At $\sigma=0$, this is the change of frame by $g$, which equals $h_1^{\#k}$ on $D^n_{\north}$ and $h_2$ on $D^n_{\rm equator}$.
Hence, the new frame agrees with the trivialization outside $D^{n+1}_{\north}$, and together they form a single trivialization of $P_G$ over $C\setminus D^{n+1}_{\rm in}$.
This trivialization does not extend over $D^{n+1}_{\rm in}$ in general, since $g'$ need not extend.
In this trivialization, $A_G$ is a one-form on all of $C\setminus D^{n+1}_{\rm in}$, and $(1-s)A_G$ with $s\in[0,1]$ is a family of connections on $P_G$.
In the original northern frame, it gives the homotopy
\begin{equation}
\label{eq:bordism-homotopy}
A_{C,s}=(1-s)A_R+s\,g_\sigma\,\d g_\sigma^{-1}
\qquad\text{on }D^{n+1}_{\north}\setminus D^{n+1}_{\rm in}.
\end{equation}
This is the analog of \eqref{eq:disk-homotopy}, with $h_1$ replaced by $g_\sigma$.
With the function $s(r)$ of Section~\ref{sec:disk-bundle-sphere-base}, we define the gauge field on the collar by \eqref{eq:disk-deformed} and \eqref{eq:disk-deformed-frame-R} with $A_{G,s}$ replaced by $A_{C,s}$:
\begin{align}
\label{eq:bordism-outer}
A&=u\,\d u^{-1}+uA^\infty u^{-1},\\
\label{eq:bordism-outer-frame-R}
A^\infty&=\frac{r^2A_{C,s(r)}+\rho^2\,u^{-1}\d u}{r^2+\rho^2},
\end{align}
where $A$ is in the frame near the instanton core, which covers the whole fiber, and $A^\infty$ is in the original northern frame away from the core.
Near $r=0$, where $s=0$, this is the instanton field \eqref{eq:disk-A1-horizontal}, and $g_\sigma$ does not enter.
As $r\to\infty$, $A$ becomes pure gauge, $A=t\,\d t^{-1}$ with $t=ug_\sigma$.

Near $S^n_{\rm in}$, we have $g_\sigma=g'$, which commutes with $\SU(2)_1$, and hence with $u$ and with $A_R$, regarded as $\su(2)_1$-valued.
We call the frame of $A$ in \eqref{eq:bordism-outer} the \emph{outer} frame, and the frame obtained by changing it by $g'$ the \emph{inner} frame.
Since $A_{C,s}=(1-s)A_R+s\,g'\,\d g'^{-1}$ near $S^n_{\rm in}$, the gauge field \eqref{eq:bordism-outer} in the inner frame is
\begin{equation}
\label{eq:bordism-4}
\begin{aligned}
A^{g'}&=g'^{-1}Ag'+g'^{-1}\d g'\\
&=\frac{r^2}{r^2+\rho^2}\left(u\,\d u^{-1}+(1-s)\,uA_Ru^{-1}\right)
+\left(1-\frac{r^2s}{r^2+\rho^2}\right)g'^{-1}\d g',
\end{aligned}
\end{equation}
with $s=s(r)$.
The coefficient of $g'^{-1}\d g'$ equals one near the instanton core and vanishes at fiber infinity.

The map $g'$ enters \eqref{eq:bordism-4} only through the one-form $g'^{-1}\d g'$.
Since an extension of this one-form does not require an extension of the map $g'$, a nonzero $[g']$ creates no obstruction.
We extend $g'^{-1}\d g'$ to a $G'$ connection $A_{G'}$ on $D^{n+1}_{\rm in}$, for example by carrying it radially inward and multiplying it by a function that equals one near $S^n_{\rm in}$ and vanishes near the center of $D^{n+1}_{\rm in}$.
The connection $A_{G'}$ generally has curvature.
In the inner frame on $D^{n+1}_{\rm in}$, we define
\begin{equation}
\label{eq:bordism-5}
A^{g'}=\frac{r^2}{r^2+\rho^2}\left(u\,\d u^{-1}+(1-s)\,uA_Ru^{-1}\right)
+\left(1-\frac{r^2s}{r^2+\rho^2}\right)A_{G'}.
\end{equation}
At fiber infinity, $A^{g'}\to u\,\d u^{-1}$, which is pure gauge.
The transition function $g'$ between the inner and outer frames on $S^n_{\rm in}$ specifies the gauge bundle seen by the instanton zero modes, as we will see in Section~\ref{sec:reduction-zero-modes}.

The gauge field is thus pure gauge at fiber infinity, i.e.~on $Y_0$.
To see that the trivializations on the three regions fit together, we write the gauge field at fiber infinity as $A=t\,\d t^{-1}$, where $A$ is in the frame of \eqref{eq:disk-deformed} and \eqref{eq:bordism-outer}, as in Section~\ref{sec:disk-bundle-sphere-base}.
We have
\begin{equation}
t=\begin{cases}
u_{\north}h_1^{\#k} & \text{on the northern part of } C\setminus D^{n+1}_{\north},\\
u_{\south} & \text{on the southern part of } C\setminus D^{n+1}_{\north},\\
u_{\north}g_\sigma & \text{on the collar } D^{n+1}_{\north}\setminus D^{n+1}_{\rm in},
\end{cases}
\end{equation}
and $t=u_{\north}$ in the inner frame on $D^{n+1}_{\rm in}$.
The first two expressions agree on the equator, as on $Y$.
The third agrees with them on $\partial D^{n+1}_{\north}$, since $g=h_1^{\#k}$ on $D^n_{\north}$ and $g=h_2$ on $D^n_{\rm equator}$, where $u_{\south}=u_{\north}h_2$.
On $S^n_{\rm in}$, the inner frame is the outer frame changed by $g'$, so $t=u_{\north}$ in the inner frame corresponds to $t=g'u_{\north}=u_{\north}g'$ in the outer frame, which agrees with the third expression at $\sigma=1$.
On the copies of $X$ in $\partial Y_0$, this trivialization is the one of Section~\ref{sec:disk-bundle-sphere-base}.

Finally, $Z$ admits a twisted string structure that restricts to the required structures on its boundary.
The base $C$ is homotopy equivalent to a wedge of $k-1$ copies of $S^n$, and $Z$ retracts onto $C$.
Since $n>4$, $H^j(Z;\bZ)=0$ for $1\le j\le4$, so $Z$ has a unique spin structure and a unique twisted string structure.
On each copy of $Y$, it restricts to the twisted string structure used in Section~\ref{sec:disk-bundle}, which is unique for the same reason.
On $Y_0$, where the gauge field is pure gauge, it restricts to a string structure whose flux through each fiber $S^3=\partial D^4$ is the instanton number, as in \eqref{eq:review-10}.

We have thus constructed a gauge field on $Z=D(V_C)$ that agrees with the one of Section~\ref{sec:disk-bundle} on each copy of $Y$ and is pure gauge on $Y_0$, together with a twisted string structure.
Coupling the sigma model on $Z$ to $\sT$ by this gauge field gives $\sL$, whose boundary is $\sN^{\sqcup k}\cup\overline{\sY_0\times\sT}$, as required by \eqref{eq:review-2}.
This completes the construction.

\section{Reducing the APS index to the base}
\label{sec:reduction}

At large target volume, the APS index of the SQFT $\mathsf L$ is the APS index of a Dirac operator on the $(n+5)$-dimensional space $Z$. This section reduces it to a Dirac index on the base. The physical picture is dimensional reduction on the fiber. Expand a fermion on $Z$ in the zero modes of the four-dimensional instanton in the fiber. Each $\mathrm{SU}(2)_1$ doublet gives one chiral zero mode localized near the instanton, while singlets give none. The coefficient of such a zero mode is a fermion on the base $C$. Its gauge representation is determined by the commutant $G'$, and its connection comes from the gauge field constructed in Section~\ref{sec:bordism}. We take the fiber scale much smaller than the base scale, so that the instanton is pointlike as seen from the base.

For the representations appearing here, the result is

\begin{equation}
\label{eq:reduction-1}
\boxed{\quad
W(\mathsf X,\mathsf T)=\frac{1}{2k}
\operatorname{Ind}_{\mathbb C}
\left(\mathcal D^+_{S^{n+1}}\otimes\mathcal V_{g'}\right)
\pmod{\mathbb Z}.
\quad}
\end{equation}

Here $\mathcal V$ records the $G'$ representations of the instanton zero modes, with a plus sign for even internal fermion parity and a minus sign for odd parity. Such a formal difference is called a \emph{virtual representation}. An index with this coefficient means the index of the positive part minus that of the negative part. The zero modes couple to the $G'$ connection $A_{G'}$ of \eqref{eq:bordism-5}, which approaches $g'^{-1}\d g'$ at the boundary of $D^{n+1}_{\rm in}$; the resulting virtual bundle is denoted $\mathcal V_{g'}$ and is specified in Section~\ref{sec:reduction-zero-modes}.

Thus \eqref{eq:reduction-1} has a direct interpretation: count the chiral zero modes that remain after the fiber instanton reduction, include the normalization $1/2$, and divide by the $k$ original copies. The factor $1/2$ is the factor $\kappa$ of \eqref{eq:review-aps}.

In this section we explain the reduction at the level needed for the calculations of Sections~\ref{sec:sp2} and~\ref{sec:su3}, together with the physical reasons that it works.\footnote{The analytic justification may use the APS index theorem~\cite{Atiyah:1975jf} and the adiabatic-limit theorem of \cite{Dai1991}.}

\subsection{Which Ramond states contribute?}
\label{sec:reduction-ramond}

We first identify which worldsheet states contribute to the constant coefficient in \eqref{eq:review-3}. At large target volume, quantizing the Ramond zero modes of the right-moving fermions gives spinors on the target, and the supercharge acts as a Dirac operator. Internal states specify the bundles to which it is coupled. The APS index $I_{\mathsf L}(\tau)$ is therefore a series of Dirac indices with APS boundary conditions, one for each momentum sector. The factor $\kappa$ in \eqref{eq:review-aps} supplies the common normalization factor $1/2$.

When $\mathsf T$ is conformal with left and right central charges $c_L$ and $c_R$, an internal state with left and right conformal weights $h$ and $\bar h$ has $H_L=h-c_L/24$ and $H_R=\bar h-c_R/24$.
An internal state with $H_R=0$ has $P=H_L$, so it contributes $q^{h-c_L/24}$. The left-moving oscillators of the bosonic fields of the sigma model on $Z$ contribute $\eta_{\mathrm D}(\tau)^{-(d+2)}$.
Including the factor $\eta_{\mathrm D}(\tau)^{4}$ in \eqref{eq:review-3}, the leading contribution of an internal state with $H_R=0$ and weight $h$ is
\begin{equation}
\label{eq:reduction-2}
q^{\,h-(d+2+c_{L}-4)/24}
=q^{\,h-c_R/24-(d+6)/16},
\end{equation}
where we used $\nu_{\mathsf T}=2(c_R-c_L)=-22-d$.
In our examples, $\mathsf T$ is purely left-moving, so $c_R=0$, and \eqref{eq:reduction-2} is $q^0$ for $h=(d+6)/16$.

If \eqref{eq:reduction-2} has no negative power of $q$ for any internal state with $H_R=0$, as in our examples, the coefficient of $q^0$ uses only the finite-dimensional virtual representation of the Ramond states with $H_R=0$ for which \eqref{eq:reduction-2} is $q^0$,
\begin{equation}
\label{eq:reduction-3}
\mathcal R=\mathcal R_+-\mathcal R_-
\end{equation}
with $\mathcal R_+$ and $\mathcal R_-$ the even and odd Ramond states. 
Denoting the coefficient of $q^0$ in \eqref{eq:review-3} by $a_0$, we obtain

\begin{equation}
\label{eq:reduction-5}
a_0=\frac12\,\operatorname{Ind}^{\mathrm{APS}}_{\mathbb C}
(\mathcal D^+_{Z}\otimes\mathcal R).
\end{equation}

\subsection{The instanton zero modes and their gauge bundle}
\label{sec:reduction-zero-modes}

We next determine the low-energy states on a single fiber. Each fiber of $Z$ is a four-disk carrying the charge-one instanton, and the $G$-connection on its boundary $S^3$ is pure gauge in the construction of Section~\ref{sec:bordism}. As in Section~\ref{sec:disk-bundle}, we identify the interior of the disk with $\mathbb R^4$. The Dirac operators on the fibers over the points of $C$ form a family parametrized by $C$.

For the charge-one BPST instanton on $\mathbb R^4$, the four-dimensional Dirac operator coupled to the doublet has complex index one for the fiber orientation with $\int_{D^4}c(F)=1$ as in \eqref{eq:review-10}, and all of its normalizable zero modes have the same chirality. It therefore has exactly one normalizable complex zero mode, of positive chirality. A gauge singlet has no normalizable zero modes. 

The zero mode from the doublet of $\mathrm{SU}(2)_1$ is a singlet under the $\mathrm{SU}(2)_R$ subgroup of the rotation group of $\mathbb R^4$.
The zero mode is therefore unaffected by the transition functions of the $\mathrm{SU}(2)_R$ bundle specified in Section~\ref{sec:disk-bundle-sphere-base}. The lower-dimensional fermion still transforms under the commutant $G'$. If
\begin{equation}
\label{eq:reduction-6}
\mathcal R|_{\mathrm{SU}(2)_1\times G'}
=\mathbf2\otimes\mathcal V+\mathbf1\otimes\mathcal U,
\end{equation}
as virtual representations, each doublet contributes its one zero mode and each singlet contributes none. The resulting virtual representation is $\mathcal V$. As the base point varies, the zero modes for $\mathcal R_+$ and $\mathcal R_-$ form vector bundles. Its rank is zero in our examples: there are equally many zero modes from the two Ramond parities at each point. The two bundles can nevertheless be nonisomorphic and give different Dirac indices on the base. This identification relies on the doublet/singlet restriction; higher isospin of $\mathrm{SU}(2)_1$ would produce additional zero modes with nontrivial rotation representations.

Since the zero mode is localized within the instanton size $\rho$ of the core, we take the function $s(r)$ of Section~\ref{sec:disk-bundle-sphere-base} to vanish up to a radius much larger than $\rho$, so that to leading order the zero mode sees only the gauge field where $s=0$. In the region inside $D^{n+1}_{\rm in}$ of Section~\ref{sec:bordism-commutant}, the horizontal part of the gauge field \eqref{eq:bordism-5} near the instanton core is $A_{G'}$, and this is what the zero modes see. The zero modes do not see $A_R$ because they are singlets under $\mathrm{SU}(2)_R$. 

Thus the fermions on $C$, in the virtual representation $\mathcal V$, couple to the $G'$ connection $A_{G'}$ on $D^{n+1}_{\rm in}$ and to pure gauge elsewhere. At the boundary $S^n_{\rm in}$ of $D^{n+1}_{\rm in}$, $A_{G'}$ is the pure gauge $g'^{-1}\d g'$. This is the familiar situation of a localized gauge field on $\mathbb{R}^{n+1}$: the gauge field is pure gauge in the region far from the origin of $\mathbb{R}^{n+1}$.

The reduction uses the separation of scales
\begin{equation}
\label{eq:reduction-8}
\ell_s\ll \rho\ll L_{\rm fiber}\ll L_{\rm base}.
\end{equation}
Here $\ell_s$ is the string length, $\rho$ the instanton size of \eqref{eq:disk-2}, and $L_{\rm fiber}$ and $L_{\rm base}$ the typical length scales of fiber and base. The function $s(r)$ of \eqref{eq:disk-deformed-frame-R} changes from $0$ to $1$ at radii of order $L_{\rm fiber}$, which is much larger than $\rho$. 

What remains is the APS index of the base problem:
\begin{equation}
\label{eq:reduction-9}
\operatorname{Ind}^{\mathrm{APS}}_{\mathbb C}
(\mathcal D_{Z}^+\otimes\mathcal R)
=
\operatorname{Ind}^{\mathrm{APS}}_{\mathbb C}
(\mathcal D_C^+\otimes\mathcal V).
\end{equation}
Since the connection seen by the fermions on $C$ is nontrivial only on $D^{n+1}_{\rm in}$, the index \eqref{eq:reduction-9} is that of the Dirac operator on $D^{n+1}_{\rm in}$ coupled to $A_{G'}$ with the asymptotic form $g'^{-1}\d g'$.
It is convenient to rewrite it as an index on the closed sphere $S^{n+1}$.
We obtain $S^{n+1}$ by gluing to $D^{n+1}_{\rm in}$ another disk, on which the connection is pure gauge, and orient it so that $D^{n+1}_{\rm in}$ keeps its orientation.
We denote the resulting virtual bundle on $S^{n+1}$ by $\mathcal V_{g'}$ and say that it is glued by $g'$.
This gives \eqref{eq:reduction-1}.

\subsection{What to compute in a concrete example}
\label{sec:reduction-recipe}

The construction reduces the calculation to four tasks, each corresponding to one part of the geometry above:

\begin{enumerate}

\item Find the internal Ramond states for which \eqref{eq:reduction-2} is $q^0$. Their virtual representation $\mathcal R$ specifies the fermions on $Z$, of dimension $n+5$.

\item Restrict $\mathcal R$ to $\mathrm{SU}(2)_1\times G'$. The fiber zero modes give the lower-dimensional fermions on $C$, of dimension $n+1$, with virtual representation $\mathcal V$.

\item Determine the bundle from the transition function $h$ of $X$: find $k$ such that $k[h]=0$, the relative class $[h_1]$ such that $\partial [h_1]=[h]$, a lift $[g]$ of $k[h_1]$, and a representative $g'$ in $G'$ such that $[g']=[g]$. These data fix the gauge bundle of the lower-dimensional fermions.

\item Compute the complex Dirac index of $\mathcal V_{g'}$ on $S^{n+1}$, divide by two to obtain the normalized SQFT index, and divide by $k$ as in \eqref{eq:reduction-1}.

\end{enumerate}

Sections~\ref{sec:sp2} and~\ref{sec:su3} carry this out for the two targets.

\section{The \texorpdfstring{$\mathrm{Sp}(2)$}{Sp(2)} calculation}
\label{sec:sp2}

We now compute $W(\Sp(2)_{H=1},\sT_{16,16})$.
Here $\Sp(2)_{H=1}$ is the sigma model on the group manifold $\Sp(2)$ with unit $H$-flux, which is an $S^3$ bundle over $S^7$, and $\sT_{16,16}$ is the chiral fermionic CFT of central charge $16$ with current algebra $\widehat{\so}(16)_{1}\times\widehat{\so}(16)_{1}$.
This pairing is related to the $\bZ_{3}$-valued discrete topological term of the non-supersymmetric $\so(16)\oplus\so(16)$ heterotic string.

\subsection{The internal theory and the relevant representation}
\label{sec:sp2-internal}

The theory $\sT_{16,16}$ is an extension of the current algebra $\widehat{\so}(16)_{1}\times\widehat{\so}(16)_{1}$ (see e.g. \cite{BoyleSmith:2023xkd}): its Hilbert spaces contain the combinations of current-algebra modules given below.
Specifying only the current algebra would not specify these combinations or their fermion parities.

Each $\widehat{\so}(16)_{1}$ factor has four integrable highest-weight modules, denoted $O,V,S,C$.
The states of lowest conformal weight in each module form a representation of the finite-dimensional algebra $\so(16)$: the singlet, the vector $\mathbf{16}$, or one of the two chiral spinors $\mathbf{128}_+$ and $\mathbf{128}_-$.
A pair such as $(V,V)$ means the tensor product of modules in the two factors of $\widehat{\so}(16)_{1}\times\widehat{\so}(16)_{1}$.
We choose fermion parity in the Ramond sector so that
\begin{equation}
\label{eq:sp2-1}
\cH_{\mathrm R,+}=(V,V)\oplus(C,C),
\qquad
\cH_{\mathrm R,-}=(S,O)\oplus(O,S).
\end{equation}
For completeness, in the Neveu--Schwarz sector the two parity sectors are
$\cH_{\mathrm{NS},+}=(O,O)\oplus(S,S)$ and
$\cH_{\mathrm{NS},-}=(C,V)\oplus(V,C)$.

The conformal weights are $h_O=0$, $h_V=1/2$, and $h_S=h_C=1$.
Weights add in a tensor product.
Thus, $(V,V)$ begins at weight one, $(C,C)$ at weight two, and each of $(S,O)$ and $(O,S)$ at weight one.
For $d=10$ and $c_R=0$, \eqref{eq:reduction-2} is $q^{h-1}$, so only the states with weight one contribute to the index.
Thus, the virtual representation is given by
\begin{equation}
\label{eq:sp2-2}
\cR=(\mathbf{16},\mathbf{16})
-(\mathbf{128}_+,\mathbf1)-(\mathbf1,\mathbf{128}_+).
\end{equation}
In particular, $\dim\cR=256-128-128=0$.\footnote{The full elliptic genus of this theory, with $\nu=-32$, vanishes. At zero chemical potential its signed Ramond count is $v^2+s^2-2so$, where $o,v=(\theta_3^8\pm\theta_4^8)/2\eta_{\mathrm D}^8$ and $s=\theta_2^8/2\eta_{\mathrm D}^8$ are the characters of the vacuum, vector and spinor modules of $\mathfrak{so}(16)_1$, and this vanishes by Jacobi's identity $\theta_3^4=\theta_2^4+\theta_4^4$. There is no mod-2 elliptic genus in this degree.}
The subscript on $\mathbf{128}_+$ labels gauge-spinor chirality; the minus signs in \eqref{eq:sp2-2} come instead from Ramond fermion parity.
Equation \eqref{eq:sp2-1} fixes that parity convention.
Reversing all Ramond parities negates the pairing.
This operation is also described as tensoring with the Arf theory, an invertible two-dimensional spin theory whose Ramond state has odd parity.

\subsection{Homotopy classes and normalization of the Dirac index}
\label{sec:sp2-homotopy}

We first identify the data of the general construction of Sections~\ref{sec:disk-bundle} and~\ref{sec:bordism} in this example: the sphere bundle $X$ and the number $k$ of copies, the group $G$ with its subgroups $\mathrm{SU}(2)_1$ and $G'$, and the maps $h_1$, $g$ and $g'$.

The target space $X=\mathrm{Sp}(2)$ is the total space of the principal bundle
\begin{equation}
\label{eq:sp2-principal}
\mathrm{Sp}(1)\to\mathrm{Sp}(2)\xrightarrow{\mathrm{pr}}S^{7}=\mathrm{Sp}(2)/\mathrm{Sp}(1),
\end{equation}
where $\mathrm{Sp}(1)\simeq\mathrm{SU}(2)$ is embedded as $q\mapsto\mathrm{diag}(1,q)$.
Its image is the stabilizer of $e_1=(1,0)\in\bH^2$, and we identify $S^{7}$ with the unit sphere in $\bH^2$ by $\mathrm{pr}(A)=Ae_{1}$.
In the notation of Section~\ref{sec:disk-bundle-general-base}, the base is $B=S^7$, so that $n=7$ and $d=10$.
The four-plane bundle $V$ is associated with \eqref{eq:sp2-principal} through the action $y\mapsto yq^{-1}$ of $q\in\Sp(1)$ on $\bH\simeq\bR^4$:
\begin{equation}
\label{eq:sp2-V}
V=\mathrm{Sp}(2)\times_{\mathrm{Sp}(1)}\mathbb H\longrightarrow S^7.
\end{equation}
This action is the rotation by $\mathrm{SU}(2)_R$ in \eqref{eq:disk-rotation}, and $\mathrm{SU}(2)_L$ acts trivially, so the second assumption of Section~\ref{sec:disk-bundle-sphere-base} holds.
The sphere bundle $X=S(V)$ is $\mathrm{Sp}(2)$ itself, since $\mathrm{Sp}(1)$ acts simply transitively on the unit quaternions.

We now determine the transition function of $V$ on the equator $S^6$.
Take local sections $f_{\north}$ and $f_{\south}$ of \eqref{eq:sp2-principal} over $D^7_{\north}$ and $D^7_{\south}$.
On $S^6$, they are related by a map $h:S^6\to\mathrm{Sp}(1)$ as
\begin{equation}
\label{eq:sp2-sections}
f_{\south}=f_{\north}\begin{pmatrix}1&\\&h\end{pmatrix}.
\end{equation}
The sections determine local trivializations of $V$, and $h$ is their transition function, as we now check.
Over $D^7_{\south}$, the trivialization is
\begin{equation}
\varphi_{\south}:D^7_{\south}\times\mathbb H\ni(x,y_{\south})\mapsto[f_{\south}(x),y_{\south}]\in V|_{D^7_{\south}}.
\end{equation}
For each $A\in\mathrm{Sp}(2)$ with $\mathrm{pr}(A)\in D^7_{\south}$, there is a unique $q_A\in\mathrm{Sp}(1)$ such that
\begin{equation}
\label{eq:sp2-inverse-q}
f_{\south}(\mathrm{pr}(A))\begin{pmatrix}1&\\&q_A\end{pmatrix}=A,
\end{equation}
and the inverse is
\begin{equation}
\varphi_{\south}^{-1}:V|_{D^7_{\south}}\ni[A,y]\mapsto(\mathrm{pr}(A),yq_A^{-1})\in D^7_{\south}\times\mathbb H,
\end{equation}
which is independent of the representative of $[A,y]$.
The trivialization $\varphi_{\north}$ over $D^7_{\north}$ is defined similarly.
On $S^6$, the transition map $\varphi_{\south}^{-1}\circ\varphi_{\north}:S^6\times\mathbb H\to S^6\times\mathbb H$ is
\begin{equation}
(x,y_{\south})
=\varphi_{\south}^{-1}\circ\varphi_{\north}(x,y_{\north})
=\varphi_{\south}^{-1}\bigl([f_{\north}(x),y_{\north}]\bigr)=\bigl(x,\,y_{\north}q_{f_{\north}(x)}^{-1}\bigr).
\end{equation}
Comparing \eqref{eq:sp2-inverse-q} with \eqref{eq:sp2-sections}, we find $q_{f_{\north}(x)}=h(x)^{-1}$.
Hence, $y_{\south}=y_{\north}h$, i.e.~$h$ is the transition function of $V$ in the convention of Section~\ref{sec:disk-bundle-sphere-base}.

The number $k$ of copies is the order of $[h]\in\pi_6(\mathrm{Sp}(1))$.
The homotopy exact sequence of \eqref{eq:sp2-principal} contains
\begin{equation}
\label{eq:sp2-3}
\pi_7(\mathrm{Sp}(2))\simeq\mathbb Z
\to\pi_7(S^7)\simeq\mathbb Z
\xrightarrow{\ \partial\ }\pi_6(\mathrm{Sp}(1))\simeq\mathbb Z_{12}
\to\pi_6(\mathrm{Sp}(2))\simeq0,
\end{equation}
where $\pi_7(S^7)\simeq\mathbb Z$ is given by the degree.
The boundary homomorphism $\partial$ has the following geometric meaning.
Since $\pi_6(\mathrm{Sp}(2))\simeq0$, $h$ can be contracted within $\mathrm{Sp}(2)$:
\begin{equation}
\label{eq:sp2-hat-h1}
\hat h_1:D^7_{\north}\to\mathrm{Sp}(2),\qquad\hat h_1|_{S^6}=h,
\end{equation}
where $h$ is regarded as $\mathrm{Sp}(2)$-valued.
Since $\mathrm{pr}\circ\hat h_1$ maps $S^6$ to the point $e_1$, it defines a map $D^7_{\north}/S^6\to S^7$.
Its degree is the relative class $[\hat h_1]\in\pi_7(S^7)$ in the sense of Section~\ref{sec:bordism-obstruction}, and $\partial$ is the restriction to $S^6$, i.e.~$\partial[\hat h_1]=[h]$.
Here, both $D^7_{\north}/S^6$ and $S^7$ are oriented by the orientation of $B=S^7$; reversing it reverses both, so $[\hat h_1]$ does not depend on this choice.

We now choose a specific contraction and determine its class, including the sign.
Since $[h]$ is invariant under continuous deformations, we may shrink $D^7_{\south}$ to a small disk around $e_1=\mathrm{pr}(1)$, where $1$ is the unit element of $\mathrm{Sp}(2)$, and choose $f_{\south}$ close to $1$.
In the limit where $D^7_{\south}$ shrinks to the point $e_1$, we have $f_{\south}=1$, and $f_{\north}$ becomes a section over $S^7\setminus\{e_1\}$, parametrized by a disk $D^7$ whose boundary is mapped to $e_1$.
Thus, $\mathrm{pr}\circ f_{\north}$ is the collapse map $D^7\to D^7/\partial D^7\simeq S^7$ of degree one, and the boundary value of $f_{\north}$ is $h^{-1}$ by \eqref{eq:sp2-sections}.
Therefore, $f_{\north}^{-1}$ is a contraction of $h$, and we take $\hat h_1=f_{\north}^{-1}$.
Since the pointwise inverse in a topological group acts as $-1$ on homotopy classes, $[\hat h_1]=-[f_{\north}]=-1$.
As $\partial$ in \eqref{eq:sp2-3} is surjective, it maps $-1$ to a generator.
Hence, $[h]=\partial[\hat h_{1}]$ generates $\pi_{6}(\Sp(1))\simeq\bZ_{12}$, and $k=12$.

The third assumption of Section~\ref{sec:disk-bundle-sphere-base} requires a contraction $h_1$ of $h$ in the symmetry group $G$ of the internal theory.
The group structure of $X$ supplies it.
A homomorphism $\Sp(2)\to G$ that maps the subgroup $\Sp(1)$ in \eqref{eq:sp2-principal} onto $\SU(2)_1$ carries $\hat h_{1}$ to a contraction $h_{1}$ in $G$, whose relative class in $\pi_{7}(G/\SU(2)_1)$ is the image of $[\hat h_{1}]=-1$.

The symmetry group of the internal theory of Section~\ref{sec:sp2-internal} is given by
\begin{equation}
\label{eq:sp2-G}
G=\frac{\Spin(16)\times\Spin(16)}{\bZ_{2}},
\end{equation}
where $\bZ_2$ is generated by the diagonal central element that acts as $-1$ on $V$ and $C$ and as $+1$ on $O$ and $S$ in each factor.\footnote{See e.g. \cite{Fraiman:2023cpa} for the gauge group of the non-supersymmetric $\so(16)\oplus\so(16)$ heterotic string theory.}
The group $\mathrm{Sp}(2)$ is embedded into the first $\mathrm{Spin}(16)\subset G$ by its real eight-dimensional representation on $\mathbb H^2$:
\begin{equation}
\label{eq:sp2-embedding}
\mathrm{Sp}(1)\subset\mathrm{Sp}(2)\subset\mathrm{Spin}(16),
\qquad
\mathbb H\subset\mathbb H^2\simeq\mathbb R^8\subset\mathbb R^{16}.
\end{equation}
Since $\mathrm{Sp}(2)$ is simply connected, this orthogonal representation lifts to $\mathrm{Spin}(16)$.
The subgroup $\Sp(1)$ of $\Sp(2)$ acts on one quaternionic line $\bH\simeq\bR^4$ of $\bH^2$, which we call the \emph{instanton four-plane}; this $\Sp(1)$ is the root $\SU(2)_1$ required in Section~\ref{sec:disk-bundle-general-base}.
It acts on the instanton four-plane as one factor of the double cover $\Spin(4)=\SU(2)_{1}\times\SU(2)_{2}$ of its rotation group.
The commutant $G'$ of $\mathrm{SU}(2)_1$ in $G$, defined in Section~\ref{sec:bordism-commutant}, is then given by
\begin{equation}
\label{eq:sp2-commutant}
G'=\frac{\SU(2)_{2}\times\Spin(12)\times\Spin(16)}{\bZ_{2}}.
\end{equation}
Here $\SU(2)_{2}$ is the other factor of $\Spin(4)$, $\Spin(12)$ acts on the complementary twelve-plane in the first $\bR^{16}$, and the second $\Spin(16)$ is a spectator throughout.\footnote{%
The commutant of $\SU(2)_1$ in the first $\Spin(16)$ is $\SU(2)_2\times\Spin(12)$ without any quotient, since the kernel of $\SU(2)_1\times\SU(2)_2\times\Spin(12)\to\Spin(16)$ is generated by $(-1,-1,-1)$ and hence each element commuting with $\SU(2)_1$ has a unique representative with trivial $\SU(2)_1$ component.
The generator of $\bZ_2$ in \eqref{eq:sp2-G} acts as $-1$ on $\mathbf{16}$ and as $+1$ on $\mathbf{128}_+$ of the first $\Spin(16)$.
By the branching rules~\eqref{eq:sp2-branching} below, its representative with trivial $\SU(2)_1$ component acts as $-1$ on the doublet of $\SU(2)_2$ and on $\mathbf{12}$ and $\mathbf{32}_-$ of $\Spin(12)$.
Hence $G'$ is $\bigl(\SU(2)_2\times\Spin(12)\times\Spin(16)\bigr)/\bZ_2$, where $\bZ_2$ is generated by this element times the central element of the second $\Spin(16)$ acting nontrivially on $V$ and $C$.
The representations in \eqref{eq:sp2-8} are invariant under this $\bZ_2$, as they should be.
}
In particular, the embedding \eqref{eq:sp2-embedding} carries $\hat h_{1}$ to the required contraction $h_{1}$.

Let $\hat g:S^7\to\mathrm{Sp}(2)$ be the map $g$ of \eqref{eq:bordism-2} with $\hat h_1$ in place of $h_1$: it equals $\hat h_1^{\#12}$ on $D^7_{\north}$ and $h_2$, regarded as $\mathrm{Sp}(1)$-valued, on $D^7_{\rm equator}$.
Its class $[\hat g]\in\pi_{7}(\Sp(2))$ maps to $12[\hat h_1]=-12$ under the first map in \eqref{eq:sp2-3}, and hence it is a generator of $\pi_7(\Sp(2))$.
Since $g$ is $\hat g$ composed with the embedding, $[g]\in\pi_7(G)$ is the image of $[\hat g]$; here $\pi_7(G)\simeq\pi_7(\Spin(16))^{\oplus2}$ because the quotient in \eqref{eq:sp2-G} is by a finite central subgroup.

Section~\ref{sec:bordism-commutant} further requires a deformation of $g$ to a map $g'$ with values in $G'$.
Move the real eight-dimensional block $\mathbb H^2$, on which the embedded $\mathrm{Sp}(2)$ acts, to an eight-plane disjoint from the instanton four-plane, inside the complementary twelve-plane.
This is a constant conjugation in the connected group $\mathrm{Spin}(16)$, hence gives a homotopy of maps from $g$ to a map $g'$ whose image commutes with $\mathrm{SU}(2)_1$.
The map $g'$ takes values in the factor $\mathrm{Spin}(12)$ of \eqref{eq:sp2-commutant}, whose inclusion into $\mathrm{Spin}(16)$ is in the stable range and induces an isomorphism on $\pi_7$; so no information about $[g]$ is lost in passing to $g'$.

The last input is the Dirac index on $S^8$ of the bundle glued by $g'$. We first normalize the index for $\hat g$ itself. Consider the bundle on $S^8$ glued by $\hat g$ in the sense of Section~\ref{sec:reduction-zero-modes}. The fundamental representation of $\mathrm{Sp}(2)$ gives a complex rank-four bundle $E_{\hat g}$ with a quaternionic structure. The word generator alone does not fix a numerical index until we specify the representation, whether the index counts complex or quaternionic zero modes, and the orientation of $S^8$. The orientation of the base $S^7$, and with it that of $X$, has not been fixed so far. We fix it by requiring\footnote{
  Since $c_1$, $c_2$ and $c_3$ vanish on $S^8$, Newton's identity gives $\operatorname{ch}_4=-\frac16c_4$, where $c_4$ is the Euler class of the underlying real bundle with its fibers $\mathbb H^2\simeq\mathbb C^4$ oriented by the complex structure. To evaluate it, write $S^8=D^8_{\rm in}\cup D^8_{\rm out}$, where $D^8_{\rm out}$ is the disk on which the connection vanishes up to gauge transformation, and take the section of $E_{\hat g}$ that equals $e_1$ in the frame on $D^8_{\rm out}$ in which the connection vanishes. It has no zeros on $D^8_{\rm out}$, and in the inner frame on $D^8_{\rm in}$ it is $\hat g^{-1}e_1$ on $S^7_{\rm in}=\partial D^8_{\rm in}$. Extend it into $D^8_{\rm in}$ as $r\,\hat g^{-1}e_1$, where $r$ is the radial coordinate of $D^8_{\rm in}$ and $\hat g$ depends only on the angular coordinates. This section vanishes only at the origin, and its index there is the degree of $\hat g^{-1}e_1:S^7_{\rm in}\to S^7$, where $S^7$ is the unit sphere of $\mathbb C^4$ oriented as the boundary of the complex-oriented unit ball. Hence $\int_{S^8}\operatorname{ch}_4(E_{\hat g})=-\frac16\deg(\hat g^{-1}e_1)$.
  Under the identification $\partial D^8_{\rm in}\simeq\partial D^8_{\north}$ by the collar, the boundary orientation of $\partial D^8_{\rm in}$ agrees on $D^7_{\north}$ with the orientation of $B$. If $B$ is oriented as the unit sphere of $\mathbb H^2\simeq\mathbb C^4$, then $\deg(\hat g^{-1}e_1)=\deg(\mathrm{pr}\circ\hat g^{-1})=-12[\hat h_1]=12$ and $\int_{S^8}\operatorname{ch}_4(E_{\hat g})=-2$.
}
\begin{equation}
\label{eq:sp2-4}
\int_{S^8}\operatorname{ch}_4(E_{\hat g})
=\operatorname{Ind}_{\mathbb C}(\mathcal D^+\otimes E_{\hat g})=-2.
\end{equation}
Here the quaternionic index is $-1$ and the complex index is $-2$, because one quaternionic zero mode has complex dimension two.

Equation \eqref{eq:sp2-4} is the definition of the orientation of $X$ used in this paper. Together with the fiber orientation \eqref{eq:review-10}, the fermion parity assignment \eqref{eq:sp2-1}, $\partial Z=Y^{\sqcup k}\cup\overline{Y_0}$, and the connection \eqref{eq:bordism-5}, it determines the sign of every index below, and hence the sign in \eqref{eq:main-result}. We do not identify this orientation with an independently specified Lie-group framing of $\mathrm{Sp}(2)$. Reversing it replaces $-1/3$ by $1/3$; the order-three conclusion is unchanged.

The underlying real bundle has complexification $E_{\hat g}\oplus\overline E_{\hat g}\simeq2E_{\hat g}$, so

\begin{equation}
\label{eq:sp2-5}
\int_{S^8}\operatorname{ch}_4
\big((E_{\hat g})_{\mathbb R}\otimes\mathbb C\big)=-4.
\end{equation}
We now transfer this normalization to $g'$. Let $V_{12,g'}$ be the real rank-twelve vector bundle on $S^8$ glued by $g'$ in the vector representation of $\mathrm{Spin}(12)$.
Since $g$ and $g'$ lift to the first $\mathrm{Spin}(16)$, and the homotopy between them can be taken there, we may use its sixteen-dimensional vector representation. In this representation, the bundle glued by $g$ is $(E_{\hat g})_{\mathbb R}$ plus eight trivial directions, and the bundle glued by $g'$ is $V_{12,g'}$ plus four trivial directions.
Since $g'$ is homotopic to $g$ in $\mathrm{Spin}(16)$, these two bundles are isomorphic, and \eqref{eq:sp2-5} gives
\begin{equation}
\label{eq:sp2-6}
\int_{S^8}\operatorname{ch}_4(V_{12,g'}\otimes\mathbb C)=-4.
\end{equation}

\subsection{The computation of the invariant}
\label{sec:sp2-invariant}

By restricting the representation $\cR$ in~\eqref{eq:sp2-2} to the instanton subgroup $\SU(2)_1$ and its commutant $G'$, we determine the virtual representation $\cV$ of the eight-dimensional fermions on the base $C$.
For this purpose, we need the branching rules along the embedding of Lie algebras,
\begin{equation}
\su(2)_{1}\oplus\su(2)_{2}\oplus\so(12)\subset\so(16),
\end{equation}
induced by the decomposition of the first $\bR^{16}$ into the instanton four-plane and its complementary twelve-plane.
For the representations of the first $\so(16)$ appearing in~\eqref{eq:sp2-2}, the branching rules are
\begin{align}\label{eq:sp2-branching}
\begin{aligned}
\mathbf{16}&=(\mathbf{2},\mathbf{2},\mathbf{1})
\oplus(\mathbf{1},\mathbf{1},\mathbf{12}),\\
\mathbf{128}_{+}&=(\mathbf{2},\mathbf{1},\mathbf{32}_{+})
\oplus(\mathbf{1},\mathbf{2},\mathbf{32}_{-}).
\end{aligned}
\end{align}
Each $\mathbf2$ in the first entry supplies one fiber zero mode by Section~\ref{sec:reduction-zero-modes}, and the other entries are the gauge charges of that mode on the base.
The fermions in $(\mathbf{16},\mathbf{16})$ in~\eqref{eq:sp2-2} supply a doublet of $\su(2)_{2}$ tensored with the $\mathbf{16}$ of the second $\so(16)$, while those in $(\mathbf{128}_{+},\mathbf{1})$ supply $\mathbf{32}_{+}$ of $\so(12)$.
The fermions in $(\mathbf{1},\mathbf{128}_{+})$ are singlets of $\su(2)_{1}$ and supply no zero mode.
Thus, the coefficient of the instanton doublet in~\eqref{eq:sp2-2} is
\begin{equation}
\label{eq:sp2-8}
\cV=(\mathbf{2},\mathbf{1},\mathbf{16})
-(\mathbf{1},\mathbf{32}_{+},\mathbf{1})
\end{equation}
under $\su(2)_{2}\oplus\so(12)\oplus\so(16)$.
Since $g'$ is supported only in $\Spin(12)$, the positive part gives $2\cdot16=32$ trivial lines, while the negative part is a nontrivial spinor bundle.
Thus, we have
\begin{equation}
\label{eq:sp2-9}
\mathcal V_{g'}=\mathbf{1}^{\oplus 32}-S^+_{12,g'}.
\end{equation}

We compute the index by the index theorem $\operatorname{Ind}_{\bC}(\cD^{+}\otimes E)=\int\widehat{A}\operatorname{ch}(E)$.
Chern roots are a convenient way to expand the traces: formally diagonalize the curvature into two-form variables and apply ordinary power-series algebra.
A real orthogonal bundle gives roots in opposite pairs. Let $\pm x_{1},\ldots,\pm x_{6}$ be these roots for $V_{12}\otimes\bC$.
Then,
\begin{align}
\operatorname{ch}_{2}(V_{12}\otimes\bC)=\sum_{i} x_{i}^{2},
\qquad
\operatorname{ch}_{4}(V_{12}\otimes\bC)
=\frac1{12}\sum_{i} x_{i}^{4}.
\end{align}
The spinor weights are half-sums of the vector roots.
Summing their exponentials gives, through degree eight, the half-spin character $32\prod_{i}\cosh(x_{i}/2)$; the term distinguishing the two chiralities first occurs in degree twelve.
Expanding this character is a direct way to express the spinor index in terms of the vector-bundle normalization already fixed in \eqref{eq:sp2-6}:
\begin{align}
\label{eq:sp2-10}
\begin{aligned}
\operatorname{ch}_{4}(S^{+}_{12})
&=\frac{1}{12}\sum_{i} x_{i}^{4}
+\frac{1}{2}\sum_{i<j}x_{i}^{2}x_{j}^{2}\\
&=-2\operatorname{ch}_{4}(V_{12}\otimes\bC)
+\frac{1}{4}\operatorname{ch}_{2}(V_{12}\otimes\bC)^{2}.
\end{aligned}
\end{align}
Since every degree-four cohomology class vanishes on $S^{8}$, the square in \eqref{eq:sp2-10} integrates to zero.
Also, $\widehat{A}(TS^{8})=1$.
Consequently, only $\operatorname{ch}_{4}$ contributes to the index integral.
Using \eqref{eq:sp2-6}, the spinor bundle and the virtual bundle \eqref{eq:sp2-9} therefore give
\begin{align}\label{eq:sp2-11}
\operatorname{Ind}_{\bC}(\cD^{+}\otimes S^{+}_{12,g'})
=-2\cdot(-4)=8,\qquad
\operatorname{Ind}_{\bC}(\cD^+\otimes\cV_{g'})=-8.
\end{align}
Equations \eqref{eq:reduction-1} with $k=12$ and \eqref{eq:sp2-11} now yield
\begin{equation}
\label{eq:sp2-12}
W(\Sp(2)_{H=1},\sT_{16,16})
=\frac{1}{12}\frac{-8}{2}=-\frac{1}{3}.
\end{equation}

\section{The \texorpdfstring{$\mathrm{SU}(3)$}{SU(3)} calculation}
\label{sec:su3}

We now compute $W(\mathrm{SU}(3)_{H=1},\sT_{16})$.
Here $\SU(3)_{H=1}$ is the sigma model on the group manifold $\SU(3)$ with unit $H$-flux, which is an $S^3$ bundle over $S^5$, and $\sT_{16}$ is the chiral fermionic CFT of central charge $15$ with current algebra $\widehat{\su}(16)_1$.
This invariant was studied in connection with the non-supersymmetric heterotic 6-brane~\cite{Kaidi:2024cbx}.

\subsection{The internal theory and the relevant representation}
\label{sec:su3-internal}

The internal theory $\sT_{16}$ is an extension of the current algebra $\widehat{\su}(16)_1$.
Its integrable highest-weight modules are labeled by $\Lambda^j\mathbf{16}$, the antisymmetric tensors with $j$ fundamental indices, for $0\leq j\leq15$.
The lowest-weight states of the module $\Lambda^j\mathbf{16}$ transform in this representation and have conformal weight
\begin{equation}
\label{eq:su3-1}
h_j=\frac{j(16-j)}{32}.
\end{equation}

The fermionic theory of central charge $15$ has Ramond sectors
\begin{equation}
\label{eq:su3-2}
\cH_{\mathrm R,+}=\cH_{\Lambda^2}\oplus\cH_{\Lambda^{10}},
\qquad
\cH_{\mathrm R,-}=\cH_{\Lambda^6}\oplus\cH_{\Lambda^{14}}.
\end{equation}
The Neveu--Schwarz sectors are
$\cH_{\mathrm{NS},+}=\cH_{\Lambda^0}\oplus\cH_{\Lambda^8}$ and
$\cH_{\mathrm{NS},-}=\cH_{\Lambda^4}\oplus\cH_{\Lambda^{12}}$.
This is the fermion parity assignment in \cite[Table~1]{BoyleSmith:2023xkd}, also used in \cite[Appendix~D.2]{Kaidi:2024cbx}.\footnote{Here $\nu=-30\equiv2\pmod8$. The mod-2 elliptic genus vanishes by \cite[Appendix~D.2]{Kaidi:2024cbx}.}
The lowest weights in $\cH_{\Lambda^{2}}$ and $\cH_{\Lambda^{14}}\cong\cH_{\overline{\Lambda^{2}\mathbf{16}}}$ are $h_2=h_{14}=7/8$, whereas those in $\cH_{\Lambda^{6}}$ and $\cH_{\Lambda^{10}}$ are $h_6=h_{10}=15/8$.
For $d=8$ and $c_R=0$, \eqref{eq:reduction-2} is $q^{h-7/8}$, so only the states with weight $7/8$ contribute to the index.
Thus, the virtual representation is given by
\begin{equation}
\label{eq:su3-3}
\mathcal R=\Lambda^2\mathbf{16}
-\overline{\Lambda^2\mathbf{16}}.
\end{equation}

\subsection{Homotopy classes and normalization of the Dirac index}
\label{sec:su3-homotopy}

We first identify the data of the general construction of Sections~\ref{sec:disk-bundle} and~\ref{sec:bordism}.
The discussion proceeds as in Section~\ref{sec:sp2-homotopy}.

The target $X=\mathrm{SU}(3)$ is the total space of the principal bundle
\begin{equation}
\label{eq:su3-principal}
\mathrm{SU}(2)\longrightarrow\mathrm{SU}(3)\longrightarrow S^5=\mathrm{SU}(3)/\mathrm{SU}(2),
\end{equation}
obtained by letting $\mathrm{SU}(3)$ act on a unit vector $e_1\in\mathbb C^3$, whose stabilizer is $\mathrm{SU}(2)=\{\mathrm{diag}(1,A)\}$ in \eqref{eq:su3-principal}.
In the notation of Section~\ref{sec:disk-bundle-general-base}, the base is $B=S^5$, so $n=5$ and $d=8$.
By identifying $\bC^2\simeq\bH$ and $\SU(2)\simeq\Sp(1)$, the four-plane bundle $V$ is associated with \eqref{eq:su3-principal} through the action $y\mapsto yq^{-1}$ of $q\in\mathrm{SU}(2)$ on $\mathbb H\simeq\mathbb R^4$:
\begin{equation}
\label{eq:su3-V}
V=\SU(3)\times_{\SU(2)}\bH \longrightarrow S^{5}.
\end{equation}
The action of $\SU(2)$ on $\bH$ is the rotation by $\SU(2)_R$ in \eqref{eq:disk-rotation}, and $\SU(2)_L$ acts trivially.
Thus, the second assumption of Section~\ref{sec:disk-bundle-sphere-base} is satisfied.
The sphere bundle $X=S(V)$ of $V$ is $\SU(3)$ itself, because $\SU(2)\cong\Sp(1)$ acts simply transitively on the unit sphere of $\bH$.
The transition function of $V$ on the equator of $S^5$ is the clutching function
\begin{equation}
\label{eq:su3-h}
h:S^4\longrightarrow\SU(2)_R,
\qquad [h]\in\pi_{4}(\SU(2))\cong\bZ_{2},
\end{equation}
and the closed base sphere on which the final index is computed is $S^{n+1}=S^6$.

The number of copies $k$ is the order of the class $[h]$ in \eqref{eq:su3-h}.
As in Section~\ref{sec:sp2-homotopy}, this class is determined by the homotopy exact sequence of the fibration $S^3\to X\to S^5$ together with the known homotopy groups of the total space $\mathrm{SU}(3)$:
\begin{equation}
\label{eq:su3-4}
\pi_{5}(\SU(3))\cong\bZ
\xrightarrow{\ \times2\ }\pi_{5}(S^{5})\cong\bZ
\longrightarrow\pi_{4}(\SU(2))\cong\bZ_{2}
\longrightarrow\pi_{4}(\SU(3))\cong0.
\end{equation}
As in Section~\ref{sec:sp2-homotopy}, the sections of \eqref{eq:su3-principal} give a contraction $\hat h_1:D^5_{\north}\to\mathrm{SU}(3)$ of $h$ with $[\hat h_1]=-1\in\pi_{5}(S^5)$, so $[h]=\partial[\hat h_1]$ generates $\bZ_{2}$.
Thus, we have $k=2$.

The third assumption of Section~\ref{sec:disk-bundle-sphere-base} requires a contraction of $h$ in the symmetry group $G$ of the internal theory.
To this end, it suffices to embed $\SU(3)$ in $G$ so that its subgroup $\SU(2)$ maps onto $\SU(2)_{1}$.
The symmetry group of the internal theory of Section~\ref{sec:su3-internal} is
\begin{equation}
\label{eq:su3-G}
G=\SU(16)/\bZ_2.
\end{equation}
Here, $\bZ_2$ is generated by $-\mathbf 1$, which acts trivially on all the modules in the Neveu--Schwarz and Ramond sectors since only $\Lambda^j\mathbf{16}$ with even $j$ appear.
Since all the maps below take values in subgroups of $\SU(16)$, the gauge bundles we construct lift to $\SU(16)$-bundles.
We use this lift to describe associated bundles, including those in the defining representation $\mathbf{16}$, which is not a representation of $G$.
We then embed $\SU(3)$ in $\SU(16)$ as a three-dimensional block, so that its subgroup $\SU(2)$ becomes the upper two-dimensional block, which is the instanton subgroup $\mathrm{SU}(2)_{1}$:
\begin{equation}
\label{eq:su3-embedding}
\SU(2)_{1}\subset\SU(3)\subset\SU(16),
\qquad
\bC^{2}\subset\bC^{3}\subset\bC^{16}.
\end{equation}

The commutant $G'$ of $\mathrm{SU}(2)_1$ in $G$ consists of the matrices commuting with the upper two-dimensional block:
\begin{equation}
\label{eq:su3-commutant}
G'=\{\mathrm{diag}(\lambda\mathbf 1_2,B):\ \lambda\in\mathrm{U}(1),\ B\in\mathrm{U}(14),\ \lambda^2\det B=1\}/\langle-\mathbf 1\rangle
\supset\mathrm{SU}(14),
\end{equation}
where $\mathrm{SU}(14)$ acts on the complementary fourteen-plane; only this subgroup will be needed. The embedding carries $\hat h_1$ to the required contraction $h_1$.
We define $\hat g:S^{5}\to\SU(3)$ from $\hat h_1$ as in Section~\ref{sec:sp2-homotopy}.
Its class $[\hat g]\in\pi_{5}(\SU(3))$ maps to $2[\hat h_1]=-2$ under the first map in \eqref{eq:su3-4}, so it is a generator, and $[g]\in\pi_5(G)\simeq\pi_5(\SU(16))$ is its image.
A constant conjugation in $\SU(16)$ moves the three-dimensional block into the complementary $\SU(14)$, which commutes with $\SU(2)_1$.
This supplies the required commuting map $g'$ of Section~\ref{sec:bordism-commutant}, with values in $\SU(14)$.

Let $E$ be the rank-fourteen bundle on $S^6$ glued by $g'$ in the sense of Section~\ref{sec:reduction-zero-modes}.
The inclusion $\SU(14)\subset G$ is an isomorphism on $\pi_5$, so $g'$ still represents a generator after the block is placed in $\SU(14)$.
For the stable complex bundle obtained from such a generator, the top Chern character integrates to $\pm1$. Choose the sign so that the Dirac index is positive:\footnote{
  The sign is opposite to that in \eqref{eq:sp2-4} because $\operatorname{ch}_3=\frac12c_3$ on $S^6$, whereas $\operatorname{ch}_4=-\frac16c_4$ on $S^8$. With $B=S^5$ oriented as the unit sphere of $\mathbb C^3$, as for $B=S^7$ in Section~\ref{sec:sp2-homotopy}, the same argument gives $\int_{S^6}c_3(E)=\deg(\hat g^{-1}e_1)=-2[\hat h_1]=2$.
}

\begin{equation}
\label{eq:su3-5}
\operatorname{Ind}_{\mathbb C}(\mathcal D^+_{S^6}\otimes E)
=\int_{S^6}\operatorname{ch}_3(E)=1.
\end{equation}

\subsection{The computation of the invariant}
\label{sec:su3-invariant}

To find the coefficient of the instanton doublet in~\eqref{eq:su3-3}, we decompose the representations under $\su(2)_{1}\oplus\su(14)\subset\su(16)$; the $\mathfrak u(1)$ factor of the commutant can be ignored since $g'$ takes values in $\SU(14)$.
The fundamental representation decomposes as $\mathbf{16}=(\mathbf{2},\mathbf{1})\oplus(\mathbf{1},\mathbf{14})$, and hence
\begin{equation}
\Lambda^{2}\mathbf{16}=(\mathbf{1},\mathbf{1})\oplus(\mathbf{2},\mathbf{14})\oplus(\mathbf{1},\Lambda^{2}\mathbf{14}),
\end{equation}
where the three terms correspond to two indices in the doublet, one index in each block, and two indices in the fourteen-dimensional block.
Only the fermions in $(\mathbf{2},\mathbf{14})$ contain an instanton doublet.
Including the conjugate representation in~\eqref{eq:su3-3}, we find
\begin{equation}
\label{eq:su3-7}
\cV=\mathbf{14}-\overline{\mathbf{14}},
\qquad \cV_{g'}=E-\overline E.
\end{equation}

Complex conjugation reverses the signs of the Chern roots, and hence acts by $(-1)^j$ on $\operatorname{ch}_j$.
The six-dimensional index uses $\operatorname{ch}_3$, so $E$ and $\overline E$ have opposite indices.
In addition, $\widehat A(TS^6)=1$.
Thus, the Dirac indices are given by
\begin{equation}
\label{eq:su3-8}
\operatorname{Ind}_{\bC}(\cD^+\otimes\overline E)=-1,
\qquad
\operatorname{Ind}_{\bC}(\cD^+\otimes(E-\overline E))=2.
\end{equation}
It follows immediately from \eqref{eq:reduction-1} that
\begin{equation}
\label{eq:su3-9}
W(\SU(3)_{H=1},\sT_{16})
=\frac{1}{2}\frac{2}{2}=\frac12.
\end{equation}
Here, the sign of the orientation, the sign of the generator in \eqref{eq:su3-5}, and an overall reversal of Ramond parity all give the same element of $\bQ/\bZ$.

\section{Comparison with earlier determinations}
\label{sec:comparison}

The $\mathbb Z_3$-valued invariant computed in Section~\ref{sec:sp2} was determined in~\cite{Tachikawa:2024ucm} from the anomaly of the NS5-brane worldvolume theory, without computing the bulk phase. As explained in Section~\ref{sec:anomaly-inflow}, anomaly inflow \eqref{eq:inflow-expected-global} leads us to expect that this determination agrees with ours. In this section we show the agreement explicitly.\footnote{We follow~\cite{Tachikawa:2024ucm} in the overall direction of the argument, but not in its technical details, which we redo in the approach of this paper.} The idea comes from the observation that an NS-brane inflated to finite size becomes the instanton of Section~\ref{sec:disk-bundle}.

In~\cite{Tachikawa:2024ucm}, the NS5-branes of the $\mathfrak{so}(32)$ and $\mathfrak e_8\oplus\mathfrak e_8$ heterotic strings, whose worldvolume theories are those of Section~\ref{sec:inflow-ns5}, are compared after restricting both gauge algebras to $\mathfrak{so}(16)\oplus\mathfrak{so}(16)$ and turning off these gauge fields. The brane then couples only to its tangent and normal bundles, and the difference of the worldvolume anomaly polynomials is\footnote{We set the dynamical $\mathfrak{sp}(1)$ gauge field of the $\mathfrak{so}(32)$ brane to zero, $c_{\rm g}=0$. This removes the term $\mathcal K_4c_{\rm g}$ of $\mathcal I_8^{\mathfrak{sp}(1)}$, which the Green--Schwarz term of that brane cancels through $\mathcal L_4^{\mathfrak{so}(32)}$ in \eqref{eq:inflow-brane-four-forms}. In principle, gauge fields on the brane require care in this comparison. They are specific to each brane: the $\mathfrak{sp}(1)$ gauge field has no counterpart on the $\mathfrak e_8\oplus\mathfrak e_8$ brane. It is therefore not clear a priori that the difference of the results for the two branes, restricted to $\mathfrak{so}(16)\oplus\mathfrak{so}(16)$, agrees with the anomaly of the NS5-brane of the non-supersymmetric $\mathfrak{so}(16)\oplus\mathfrak{so}(16)$ heterotic string theory.}
\begin{equation}
\label{eq:comparison-brane-difference}
\Delta\mathcal I_8:=\mathcal I_8^{\mathfrak{sp}(1)}-\mathcal I_8^{\mathrm{E\text{-}str}}
=\mathcal K_4\,\frac{c(R)}{3},
\qquad
\mathcal K_4=-\lambda(T\Sigma)-2c(R),
\end{equation}
where $\mathcal K_4$ is the worldvolume Bianchi four-form \eqref{eq:inflow-bianchi} at $c=0$. Here $\Delta\mathcal I_8$ and $\mathcal K_4$ are evaluated with the $\mathfrak{so}(16)\oplus\mathfrak{so}(16)$ gauge fields turned off; we use the same symbols when these gauge fields are kept, as in \eqref{eq:comparison-delta-I8} below. Since $\mathcal K_4=\d(H|_\Sigma)$, the worldvolume Green--Schwarz term cancels $\Delta\mathcal I_8$ perturbatively, as in \eqref{eq:inflow-completed-brane}. A global anomaly remains: the anomaly theory of the difference, including this term, assigns a phase $\exp(2\pi\i\mathcal A)$ as in \eqref{eq:inflow-expected-global} to each closed seven-dimensional base $B$. For $B=S^7$ with the $\mathrm{SU}(2)_R$ bundle of $\mathrm{Sp}(2)\to S^7$, this phase was found to have order three, by extending the structure group $\mathrm{SU}(2)_R$ to $\mathrm{Sp}(2)$ and evaluating a Dirac index on $S^8$. Section~4 of~\cite{Tachikawa:2024ucm} also proposed to obtain the corresponding bulk coupling from an instanton in each fiber, without evaluating it; Sections~\ref{sec:disk-bundle}--\ref{sec:sp2} carry out this evaluation.

\paragraph{The bulk phase.} On the bulk side, consider the phases $\Phi^{\mathfrak{so}(32)}(X)$ and $\Phi^{\mathfrak e_8\oplus\mathfrak e_8}(X)$ of \eqref{eq:inflow-spectral-gs} for the $\mathfrak{so}(32)$ and $\mathfrak e_8\oplus\mathfrak e_8$ strings on a general $X=\partial Y$, where the gauge fields are restricted to $\mathfrak{so}(16)\oplus\mathfrak{so}(16)$. Let us see each of the factors appearing in \eqref{eq:inflow-spectral-gs}. The massless fermions determine $\eta(Y)$ directly. In the $\mathfrak{so}(32)$ string minus the $\mathfrak e_8\oplus\mathfrak e_8$ string, the gravitino and the dilatino cancel, and the gauginos differ by the virtual representation $\mathcal R$ of \eqref{eq:sp2-2}, since $\mathbf{496}$ and $(\mathbf{248},\mathbf1)+(\mathbf1,\mathbf{248})$ share $(\mathbf{120},\mathbf1)+(\mathbf1,\mathbf{120})$. This $\mathcal R$ is the massless fermion content of the non-supersymmetric $\mathfrak{so}(16)\oplus\mathfrak{so}(16)$ string, with positive chirality in $(\mathbf{16},\mathbf{16})$ and negative chirality in $(\mathbf{128}_+,\mathbf1)+(\mathbf1,\mathbf{128}_+)$. Next, in both the $\mathfrak{so}(32)$ and $\mathfrak e_8\oplus\mathfrak e_8$ strings, the level-one form restricts to $c=a+a'$, where $a$ and $a'$ are the level-one forms of the two $\mathfrak{so}(16)$ factors. This is also $c$ of the $\mathfrak{so}(16)\oplus\mathfrak{so}(16)$ string. The three strings thus share the Bianchi identity \eqref{eq:inflow-bulk-bianchi} and hence $H$. The anomaly polynomial $\mathcal I_{12}$ of the fermions in $\mathcal R$ is the difference of \eqref{eq:inflow-factorization} for the two supersymmetric strings,
\begin{equation}
\label{eq:comparison-bulk-I12}
\begin{aligned}
\mathcal I_{12}&=\bigl(a+a'-\lambda(TX)\bigr)\mathcal J_8,\\
\mathcal J_8&=\mathcal J_8^{\mathfrak{so}(32)}-\mathcal J_8^{\mathfrak e_8\oplus\mathfrak e_8}\\
&=\frac{a^2+aa'+a'^2-p_2(V_{16})-p_2(V'_{16})}{6},
\end{aligned}
\end{equation}
where $\mathcal J_8^{\mathfrak{so}(32)}$ and $\mathcal J_8^{\mathfrak e_8\oplus\mathfrak e_8}$ are restricted to $\mathfrak{so}(16)\oplus\mathfrak{so}(16)$, and $V_{16}$ and $V'_{16}$ are the vector bundles of the two $\mathfrak{so}(16)$ factors, so that $a=p_1(V_{16})/2$ and $a'=p_1(V'_{16})/2$. Thus $\mathcal J_8$ is the eight-form of the $\mathfrak{so}(16)\oplus\mathfrak{so}(16)$ string. Hence the phase \eqref{eq:inflow-spectral-gs} of the $\mathfrak{so}(16)\oplus\mathfrak{so}(16)$ string is
\begin{equation}
\label{eq:comparison-bulk-phases}
\Phi(X)
=\frac{\Phi^{\mathfrak{so}(32)}(X)}{\Phi^{\mathfrak e_8\oplus\mathfrak e_8}(X)}.
\end{equation}

\paragraph{The expectation.} In particular, for $X=S(V)=\mathrm{Sp}(2)$ over $B=S^7$ with no gauge field on $X$, the anomaly inflow \eqref{eq:inflow-expected-global} for $\Phi^{\mathfrak{so}(32)}(X)$ and $\Phi^{\mathfrak e_8\oplus\mathfrak e_8}(X)$, together with \eqref{eq:inflow-physical-phase} for $\Phi(X)$ in \eqref{eq:comparison-bulk-phases}, predicts
\begin{equation}
\label{eq:comparison-inflow}
\exp(2\pi\i\mathcal A)
=\exp\bigl(-2\pi\i W(\mathrm{Sp}(2)_{H=1},\mathsf T_{16,16})\bigr),
\end{equation}
where $\mathcal A:=\mathcal A_\text{brane}^{\mathfrak{so}(32)}(B)-\mathcal A_\text{brane}^{\mathfrak e_8\oplus\mathfrak e_8}(B)$ is the difference of the anomalies of the NS5-branes of the two strings, each including its worldvolume Green--Schwarz term $-H\mathcal L_4$ of \eqref{eq:inflow-completed-brane}. Both sides of \eqref{eq:comparison-inflow} are amplitudes on the same oriented $B$, and no convention enters beyond those fixed in Section~\ref{sec:anomaly-inflow}. The perturbative part of this relation was verified in Section~\ref{sec:inflow-ns5}; its global part is what the two calculations compare.

\paragraph{The instanton as an inflated NS5-brane.} In the inflow picture, the fibers of $X=S(V)$ are spheres linking an NS5-brane. In our construction, they bound the disks of $Y=D(V)$, each carrying an instanton that supplies the same $H$-flux. The two descriptions are the ends of one family: as the instanton size $\rho\to0$, the instanton becomes an NS-brane on the zero section, \eqref{eq:disk-c-limit}. 

For the supersymmetric strings, $\rho$ is a modulus on the Higgs branch of the worldvolume theory, along which anomalies are matched. At finite $\rho$, the anomaly of the brane is therefore carried by the zero modes of the ten-dimensional fermions on the instanton, which are the lower-dimensional fermions of Section~\ref{sec:reduction-zero-modes}. In this matching, the flavor background of the brane is the gauge field at fiber infinity, which is $A_R$ in the frame of \eqref{eq:disk-4}, and not its value $A_L$ at the core in the frame of \eqref{eq:disk-3}, since on the Higgs branch $\mathrm{SU}(2)_1$ is locked to $\mathrm{SU}(2)_R$ (Section~\ref{sec:disk-bundle-general-base}). Similarly, on the Higgs branch of the $\mathfrak{so}(32)$ brane, the $\mathfrak{sp}(1)$ gauge field is locked to the factor $\mathrm{SU}(2)_2$ of the commutant \eqref{eq:sp2-commutant} of $\mathrm{SU}(2)_1$, so that $c_{\rm g}=c(\mathrm{SU}(2)_2)$. We keep $\mathrm{SU}(2)_2$ trivial, as in our construction, which corresponds to $c_{\rm g}=0$ in \eqref{eq:comparison-brane-difference}.\footnote{With an $\mathrm{SU}(2)_2$ background included in the flavor bundles, \eqref{eq:comparison-local} below holds with $c_{\rm g}=c(\mathrm{SU}(2)_2)$ in $\Delta\mathcal L_4$ of \eqref{eq:comparison-delta-I8}.} 

With these identifications, a direct calculation for the instanton of Section~\ref{sec:disk-bundle-general-base}, with an arbitrary $\mathrm{Spin}(12)\times\mathrm{Spin}(16)$ background pulled back from the base, gives the relation of characteristic classes
\begin{equation}
\label{eq:comparison-local}
\begin{aligned}
\mathcal I_{12}&=\delta_\Sigma\,\Delta\mathcal I_8+\mathcal I_{12}^{(0)},\\
\Delta\mathcal I_8&=\mathcal I_8^{\mathfrak{sp}(1)}-\mathcal I_8^{\mathrm{E\text{-}str}}.
\end{aligned}
\end{equation}
Here $\mathcal I_{12}$ is the anomaly polynomial \eqref{eq:comparison-bulk-I12} of the ten-dimensional fermions in $\mathcal R$, $\delta_\Sigma$ is the Thom form \eqref{eq:inflow-source} of $V$ on the zero section $\Sigma$, and $\mathcal I_{12}^{(0)}$ is $\mathcal I_{12}$ with the instanton removed, so that the gauge field is the flavor background everywhere. In $\Delta\mathcal I_8$ here, unlike in \eqref{eq:comparison-brane-difference}, the $\mathfrak{so}(16)\oplus\mathfrak{so}(16)$ gauge fields are kept as the flavor background of the brane. Subtracting \eqref{eq:inflow-brane-quotient} for the two theories gives
\begin{equation}
\label{eq:comparison-delta-I8}
\begin{aligned}
\Delta\mathcal I_8
&=\mathcal J_8+\mathcal K_4\,\Delta\mathcal L_4,\\
\mathcal K_4
&=a+a'-\lambda(T\Sigma)-2c(R),\\
\Delta\mathcal L_4
&=\mathcal L_4^{\mathfrak{so}(32)}-\mathcal L_4^{\mathfrak e_8\oplus\mathfrak e_8}
=c_{\rm g}+\frac{c(R)}{3}-\frac{2a-a'}{6},
\end{aligned}
\end{equation}
where $a$ and $a'$ are those of the flavor bundles, with $\mathrm{SU}(2)_1$ in the first $\mathfrak{so}(16)$, $\mathcal L_4^{\mathfrak{so}(32)}$ and $\mathcal L_4^{\mathfrak e_8\oplus\mathfrak e_8}$ are \eqref{eq:inflow-brane-four-forms}, and $\mathcal J_8$ is \eqref{eq:comparison-bulk-I12} evaluated on the flavor bundles. By \eqref{eq:comparison-bulk-I12}, $\mathcal J_8=0$ at zero flavor background, and hence $\mathcal I_{12}^{(0)}=0$ and, at $c_{\rm g}=0$, $\Delta\mathcal I_8$ reduces to \eqref{eq:comparison-brane-difference}.

In the instanton background, the flavor background is the gauge field at fiber infinity. In the vector representation of the first $\mathfrak{so}(16)$, its bundle is the sum of the instanton four-plane, on which $\mathrm{SU}(2)_1$ acts, and the complementary twelve-plane of the commutant bundle. At fiber infinity, the bundle of the four-plane is induced from the $\mathrm{SU}(2)_R$ bundle through $\mathrm{SU}(2)_R\simeq\mathrm{SU}(2)_1$ by \eqref{eq:disk-4}, while that of the twelve-plane is the same as at the core. Written in terms of the commutant bundle, $\Delta\mathcal I_8$ is the index density of the zero modes of Section~\ref{sec:reduction-zero-modes},
\begin{equation}
\label{eq:comparison-zero-modes}
\Delta\mathcal I_8=\frac12\bigl[\widehat A(T\Sigma)\operatorname{ch}(\mathcal V)\bigr]_8,
\end{equation}
where $\mathcal V$ of \eqref{eq:sp2-8} is coupled to the commutant bundle, and the factor $\frac12$ for Majorana--Weyl fermions is the one in \eqref{eq:reduction-1}. The instanton thus adds to the bulk anomaly exactly the anomaly of the brane. 

A trivial commutant bundle does not give zero flavor background, because the instanton four-plane still carries the bundle induced from $\mathrm{SU}(2)_R$, so that $a=c(R)$ and $a'=0$ in \eqref{eq:comparison-delta-I8}. As we will see below, zero flavor background instead requires a commutant bundle whose twelve-plane is stably the inverse of the instanton four-plane, so that the whole flavor bundle is trivial, and for such a bundle the right-hand side of \eqref{eq:comparison-zero-modes} is equal to \eqref{eq:comparison-brane-difference}.

\paragraph{The brane anomaly.} Equation \eqref{eq:comparison-local} fixes only the perturbative part, while $\mathcal A$ is a global phase. We evaluate it in the same way as $W$ in Sections~\ref{sec:reduction} and~\ref{sec:sp2}, following \cite[Section~2]{Tachikawa:2025flw}: for twelve copies of $B$, the real-valued exponent of the phase is an APS index over a bordism, and this equality of real numbers is divided by twelve. 

Since $\Omega^{\mathrm{spin}}_7(B\mathrm{SU}(2))=0$, the anomaly theory of the difference without the worldvolume Green--Schwarz term is determined by its anomaly polynomial $\Delta\mathcal I_8$ of \eqref{eq:comparison-brane-difference}~\cite{Tachikawa:2024ucm}. We therefore write it with auxiliary fermions: six-dimensional complex Weyl fermions in the virtual representation $\mathcal R_{\rm aux}=8\cdot\mathbf2-\mathbf3-13\cdot\mathbf1$ of $\mathrm{SU}(2)_R$, with the doublet $\mathbf2$ and the triplet $\mathbf3$, whose anomaly polynomial is $[\widehat A(T\Sigma)\operatorname{ch}(\mathcal R_{\rm aux})]_8=\Delta\mathcal I_8$. Let $\eta_{\rm aux}(B)$ be the $\eta$ invariant of their Dirac operator on $B$, normalized as in \eqref{eq:inflow-aps}. The worldvolume Green--Schwarz term of the difference is $-H\,c(R)/3$, given by $\Delta\mathcal L_4$ of \eqref{eq:comparison-delta-I8} at $c_{\rm g}=0$ and zero flavor background. Then the phase, whose perturbative part is \eqref{eq:inflow-completed-brane}, is
\begin{equation}
\label{eq:comparison-brane-phase}
\exp(2\pi\i\mathcal A)
=\exp\Bigl(-2\pi\i\,\eta_{\rm aux}(B)-2\pi\i\int_BH\,\frac{c(R)}{3}\Bigr),
\end{equation}
where $\d H=\mathcal K_4$ on $B$. Twelve copies of $B$ bound the base $C$ of Section~\ref{sec:bordism}, $\partial C=\bigsqcup_{12}B$. The $\mathrm{SU}(2)_R$ bundle extends over $C$, and so does $H$, because $H^3(C;\mathbb Z)=H^4(C;\mathbb Z)=0$. The $\eta$ invariant of twelve identical copies of $B$ is twelve times that of one, and $\Delta\mathcal I_8=\d(H\,c(R)/3)$ on $C$. The APS index theorem \eqref{eq:inflow-aps} on $C$ and Stokes' theorem therefore give
\begin{equation}
\label{eq:comparison-brane-aps}
\begin{aligned}
\operatorname{Ind}^{\mathrm{APS}}_{\mathbb C}(\mathcal D_C^+\otimes\mathcal R_{\rm aux})
&=\int_C\Delta\mathcal I_8+12\,\eta_{\rm aux}(B)\\
&=12\Bigl(\eta_{\rm aux}(B)+\int_BH\,\frac{c(R)}{3}\Bigr).
\end{aligned}
\end{equation}
Since this is an equality of real numbers, not only modulo integers, we may divide it by twelve before exponentiating, and \eqref{eq:comparison-brane-phase} becomes
\begin{equation}
\label{eq:comparison-brane-phase-aps}
\exp(2\pi\i\mathcal A)
=\exp\Bigl(-\frac{2\pi\i}{12}
\operatorname{Ind}^{\mathrm{APS}}_{\mathbb C}(\mathcal D_C^+\otimes\mathcal R_{\rm aux})\Bigr).
\end{equation}

To evaluate this index, we follow~\cite{Tachikawa:2024ucm} and extend the structure group $\mathrm{SU}(2)_R$ to $\mathrm{Sp}(2)$ through $\mathrm{SU}(2)_R\simeq\mathrm{Sp}(1)\subset\mathrm{Sp}(2)$ of \eqref{eq:sp2-principal}, and $\mathcal R_{\rm aux}$ to the virtual representation $\widetilde{\mathcal R}_{\rm aux}=10\cdot\mathbf4-\mathbf{10}-30\cdot\mathbf1$ of $\mathrm{Sp}(2)$, with the fundamental $\mathbf4$ and the adjoint $\mathbf{10}$, which restricts to $\mathcal R_{\rm aux}$. On each boundary copy of $B$, the contraction $\hat h_1$ of \eqref{eq:sp2-hat-h1} trivializes the $\mathrm{Sp}(2)$ bundle. In this trivialization we deform the connection to zero along the same path on all twelve copies, in the same way as \eqref{eq:disk-homotopy}, which changes the APS index by a multiple of twelve. We then cap the boundary spheres with disks carrying the trivial connection, which contribute nothing because the positive and negative parts of $\widetilde{\mathcal R}_{\rm aux}$ have the same rank. This closes $C$ into a sphere $S^8$ with an $\mathrm{Sp}(2)$ bundle, and the APS index becomes $\operatorname{Ind}_{\mathbb C}(\mathcal D^+_{S^8}\otimes\widetilde{\mathcal R}_{\rm aux})$ modulo twelve. This is the index on $S^8$ evaluated in~\cite{Tachikawa:2024ucm}, where a single copy of $B$ is used and the factor twelve comes from the multiplication by twelve in \eqref{eq:sp2-3}.

\paragraph{The indices on $S^8$.} The bulk phase is $\exp(-2\pi\i W)$ with $W=W(\mathrm{Sp}(2)_{H=1},\mathsf T_{16,16})$ by \eqref{eq:inflow-physical-phase}. This phase depends only on $W$ modulo integers, which \eqref{eq:reduction-1} with $k=12$ gives as an index on the same oriented $S^8$, with $\mathcal V_{g'}$ built from the contraction $h_1$ given by the same $\hat h_1$ (Section~\ref{sec:sp2-homotopy}). Both sides of \eqref{eq:comparison-inflow} are therefore given by indices on $S^8$,
\begin{equation}
\label{eq:comparison-sphere-indices}
\begin{aligned}
\exp(2\pi\i\mathcal A)
&=\exp\Bigl(-\frac{2\pi\i}{12}
\operatorname{Ind}_{\mathbb C}\bigl(\mathcal D^+_{S^8}\otimes\widetilde{\mathcal R}_{\rm aux}\bigr)\Bigr),\\
\exp(-2\pi\i W)
&=\exp\Bigl(-\frac{2\pi\i}{12}\cdot\frac12
\operatorname{Ind}_{\mathbb C}\bigl(\mathcal D^+_{S^8}\otimes\mathcal V_{g'}\bigr)\Bigr),
\end{aligned}
\end{equation}
where both exponents have the sign fixed by that of $\eta$ in \eqref{eq:inflow-aps}. Thus \eqref{eq:comparison-inflow} holds if the two indices agree modulo twelve.

We now show that the two indices are in fact equal. The argument uses that the brane sees no flavor background in either calculation. In~\cite{Tachikawa:2024ucm}, this is the choice of background. In our construction, it is forced by the boundary condition: $\mathsf T$ is not coupled on $Y_0$, so the gauge field at fiber infinity must be pure gauge over all of $C$. We first derive from this condition a relation between the bundles on $S^8$ of the two calculations, and then the equality of their index densities.

The gauge bundle induced from the $\mathrm{SU}(2)_R$ bundle alone cannot satisfy the condition of no flavor background. As shown in Section~\ref{sec:bordism-obstruction}, it cannot be trivialized at fiber infinity over all of $C$ compatibly with the trivializations by $h_1$ on the twelve copies of $Y$: over the northern half-ball $D^8_{\north}$ of the cap, it keeps the transition function $g$ of \eqref{eq:bordism-2}, whose class $[g]\in\pi_7(G)$ is the obstruction.

In Section~\ref{sec:bordism-commutant}, this obstruction is removed by deforming $g$ through $g_\sigma$ into a map $g'$ with values in the commutant $G'$. Since $g'$ commutes with the instanton, it is used as the transition function between the inner and outer frames on $S^7_{\rm in}$. In the vector representation of the first $\mathrm{Spin}(16)$, $g$ acts on the eight-plane $\mathbb H^2$ of \eqref{eq:sp2-embedding}, and $g'$ is obtained from $g$ by a constant conjugation that moves $\mathbb H^2$ off the instanton four-plane (Section~\ref{sec:sp2-homotopy}). We take the image of $\mathbb H^2$ to be its orthogonal complement. Then $\mathbb R^{16}=\mathbb H^2\oplus\mathbb H^2$, with $g$ acting on the first $\mathbb H^2$ and $g'$ on the second.

Identify $S^7_{\rm in}$ with $\partial D^8_{\north}$ through the collar. For the gauge bundle induced from the $\mathrm{SU}(2)_R$ bundle alone, the flavor bundle on $S^8$ has the nontrivial transition function $g$ on the first $\mathbb H^2$. The construction of Section~\ref{sec:bordism-commutant} cancels it by inserting $g'^{-1}$ on the second $\mathbb H^2$. Since $g_\sigma g'^{-1}$ connects $g\,g'^{-1}$ to the identity, the flavor bundle glued by $g\,g'^{-1}$ is trivial. The inserted $g'^{-1}$ is the transition function between the inner and outer frames.\footnote{The transition functions $g'$ and $g'^{-1}$ differ only in whether the relation of the frames on $S^7_{\rm in}$ is read from the inside to the outside or from the outside to the inside. On $S^7_{\rm in}$, let $v_{\rm in}$ and $v_{\rm out}$ be the components of a vector of the second $\mathbb H^2$ in the inner frame, used inside $D^8_{\rm in}$, and in the outer frame, used outside it. Since the gauge field in the inner frame is $A^{g'}=g'^{-1}Ag'+g'^{-1}\d g'$ of \eqref{eq:bordism-4}, they are related by $v_{\rm out}=g'v_{\rm in}$, that is, $v_{\rm in}=g'^{-1}v_{\rm out}$. On $\partial D^8_{\north}$, the components of a vector of the first $\mathbb H^2$ in the original northern frame inside $D^8_{\north}$ and in the frame of the rest of $C$ outside it are related by $v_{\rm in}=g\,v_{\rm out}$. In this direction, the second $\mathbb H^2$ is glued by $g'^{-1}$.} Therefore the bundle $V_{8,g'}$ of the second $\mathbb H^2$ glued by $g'^{-1}$ is stably the inverse of the bundle of the first $\mathbb H^2$ glued by $g$. 

Let $\mathbb R^4$ be the orthogonal complement of the instanton four-plane in the first $\mathbb H^2$. The commutant bundle $V_{12,g'}$ of \eqref{eq:sp2-6}, from which the zero modes form $\mathcal V_{g'}$ through \eqref{eq:sp2-8}, is then $\mathbb R^4\oplus V_{8,g'}$, and it is also stably the inverse of the bundle of the first $\mathbb H^2$.

Since $g$ is the image of the map $\hat g$ of Section~\ref{sec:sp2-homotopy} with values in $\mathrm{Sp}(2)$, the bundle of the first $\mathbb H^2$ is induced through \eqref{eq:sp2-embedding} from an $\mathrm{Sp}(2)$ bundle on $S^8$ glued by $\hat g$ on $\partial D^8_{\north}$. This $\mathrm{Sp}(2)$ bundle is obtained from the $\mathrm{SU}(2)_R$ bundle on $C$ by extending the structure group through $\mathrm{SU}(2)_R\simeq\mathrm{Sp}(1)\subset\mathrm{Sp}(2)$. It is trivialized by $\hat h_1$ on each boundary copy of $B$ and is trivial on the twelve disks. Let $\widetilde E$ be the complex rank-four bundle with a quaternionic structure associated with it through the fundamental representation $\mathbf4$. Through $\mathbb C^4\simeq\mathbb H^2\simeq\mathbb R^8$, the bundle of the first $\mathbb H^2$ is then the underlying real rank-eight bundle $\widetilde E_{\mathbb R}$, and $V_{12,g'}$ is stably the inverse of $\widetilde E_{\mathbb R}$.

In terms of Pontryagin classes, this relation reads
\begin{equation}
\label{eq:comparison-stable-inverse}
p(V_{12,g'})=p(\widetilde E_{\mathbb R})^{-1}.
\end{equation}
This is the condition of zero flavor background stated below \eqref{eq:comparison-zero-modes}, with the instanton four-plane replaced by the first $\mathbb H^2$, whose bundle on $S^8$ is $\widetilde E_{\mathbb R}$. There, $\frac12[\widehat A(T\Sigma)\operatorname{ch}(\mathcal V)]_8$ of \eqref{eq:comparison-zero-modes}, which depends on the commutant bundle only through its Pontryagin classes, becomes $\Delta\mathcal I_8$ of \eqref{eq:comparison-brane-difference}, which is $[\widehat A(T\Sigma)\operatorname{ch}(\mathcal R_{\rm aux})]_8$. 

The brane calculation uses the same $\mathrm{Sp}(2)$ bundle, so that the auxiliary fermions couple to $\widetilde{\mathcal R}_{\rm aux}(\widetilde E)$. With \eqref{eq:comparison-stable-inverse}, we need
\begin{equation}
\label{eq:comparison-density-equality}
\frac12\bigl[\widehat A(TS^8)\operatorname{ch}(\mathcal V_{g'})\bigr]_8
=\bigl[\widehat A(TS^8)\operatorname{ch}\bigl(\widetilde{\mathcal R}_{\rm aux}(\widetilde E)\bigr)\bigr]_8,
\end{equation}
where the left-hand side is the index density of the zero modes and the right-hand side is that of the auxiliary fermions. This does not follow from \eqref{eq:comparison-zero-modes}, which concerns an $\mathrm{SU}(2)_R$ bundle. Indeed, $p_2(\widetilde E_{\mathbb R})$ contains the class $c_4(\widetilde E)$, which has no counterpart for $\mathrm{SU}(2)_R$, and the extension of $\mathcal R_{\rm aux}$ to $\mathrm{Sp}(2)$ is not unique. A direct calculation shows that \eqref{eq:comparison-density-equality} holds for the extension $\widetilde{\mathcal R}_{\rm aux}$. Integrating \eqref{eq:comparison-density-equality} over $S^8$ gives
\begin{equation}
\label{eq:comparison-index-equality}
\operatorname{Ind}_{\mathbb C}\bigl(\mathcal D^+_{S^8}\otimes\widetilde{\mathcal R}_{\rm aux}\bigr)
=\frac12\operatorname{Ind}_{\mathbb C}\bigl(\mathcal D^+_{S^8}\otimes\mathcal V_{g'}\bigr)
\end{equation}
and hence, by \eqref{eq:comparison-sphere-indices}, we finally get the relation \eqref{eq:comparison-inflow} predicted by anomaly inflow. This shows that the phase $\exp(2\pi\i\mathcal A)$ given by the difference of the anomalies of the two NS5-branes equals the bulk phase $\exp(-2\pi\i W)$ of the $\mathfrak{so}(16)\oplus\mathfrak{so}(16)$ string on $X=\mathrm{Sp}(2)$.

This argument involves no choice of the orientation of $B$ or of generators of homotopy groups.  The agreement is forced whenever the two calculations use the same conventions.

\section*{Acknowledgements}
We thank Yuji~Tachikawa for helpful discussions.
The work of KY and HW is supported in part by JST FOREST Program (Grant Number JPMJFR2030, Japan), 
and JSPS KAKENHI (21K03546). 

Claude Opus 5.5 (Anthropic) and GPT-6 Astra (OpenAI) were used as assistants for the calculations and for drafting the manuscript, which the authors then edited. The authors take full responsibility for the validity of all the calculations in the paper.

\bibliographystyle{ytphys}
\begingroup
\let\sb=_\let\sp=^
\bibliography{inspire,other}
\endgroup

\end{document}